\documentclass[letterpaper,twocolumn,10pt]{article}
\usepackage{usenix2019_v3}

\usepackage{amsmath}
\usepackage{booktabs}
\usepackage{hyperref}
\usepackage{caption}
\usepackage{subcaption}
\usepackage{nopageno}
\usepackage{times}
\usepackage{tikz}
\newcommand*\circled[1]{\tikz[baseline=(char.base)]{
            \node[shape=circle,draw,inner sep=0.5pt] (char) {#1};}}
\usepackage{graphicx}
\usepackage{amsmath}
\usepackage{amssymb}
\usepackage{commath}
\usepackage{multirow}
\usepackage{xspace}
\usepackage{dblfloatfix}
\newcommand{\mlduplication}{Deep-Dup\xspace}
\usepackage{bm}
\usepackage{amsmath}

\usepackage{url}

\usepackage{amsmath,amsfonts,bm}

\def\eqref#1{equation~\ref{#1}}

\def\1{\bm{1}}

\def\vt{{\bm{t}}}

\def\vx{{\bm{x}}}

\DeclareMathAlphabet{\mathsfit}{\encodingdefault}{\sfdefault}{m}{sl}
\SetMathAlphabet{\mathsfit}{bold}{\encodingdefault}{\sfdefault}{bx}{n}

\newcommand{\E}{\mathbb{E}}

\begin{document}

\date{}

\title{Deep-Dup: An Adversarial Weight \underline{Dup}lication Attack Framework to Crush Deep Neural Network in Multi-Tenant FPGA}
\author{
{\rm Adnan Siraj Rakin * \ }\\
Arizona State University \\
asrakin@asu.edu
 \and
 {\rm Yukui Luo * \ }\\
Northeastern University \\
luo.yuk@northeastern.edu 
\and 
{\rm  Xiaolin Xu  \ }\\
Northeastern University \\
x.xu@northeastern.edu
\and
 {\rm Deliang Fan  \ }\\
Arizona State University \\
dfan@asu.edu \\

 {\rm *Both Authors Contributed Equally \ }
}

\maketitle

\begin{abstract}
The wide deployment of Deep Neural Networks (DNN) in high-performance cloud computing platforms brought to light multi-tenant cloud field-programmable gate arrays (FPGA) as a popular choice of accelerator to boost performance due to its hardware reprogramming flexibility. 
Such a multi-tenant FPGA setup for DNN acceleration potentially exposes DNN interference tasks under severe threat from malicious users. This work, to the best of our knowledge, is the first to explore DNN model vulnerabilities in multi-tenant FPGAs. We propose a novel adversarial attack framework: \textit{\mlduplication}, in which the adversarial tenant can inject adversarial faults to the DNN model in the victim tenant of FPGA. Specifically, she can aggressively overload the shared power distribution system of FPGA with malicious power-plundering circuits, achieving \textit{adversarial weight duplication (AWD) hardware attack} that duplicates certain DNN weight packages during data transmission between off-chip memory and on-chip buffer, to hijack the DNN function of the victim tenant. 
Further, to identify the most vulnerable DNN weight packages for a given malicious objective, we propose a generic vulnerable weight package searching algorithm, called \emph{Progressive Differential Evolution Search (P-DES)}, which is, for the first time, adaptive to both deep learning white-box and black-box attack models. 
The proposed \mlduplication is experimentally validated in a developed multi-tenant FPGA prototype, for two popular deep learning applications, i.e., Object Detection and Image Classification. Successful attacks are demonstrated in six popular DNN architectures (e.g., YOLOv2, ResNet-50, MobileNet, etc.) on three datasets (COCO, CIFAR-10, and ImageNet). \footnote{Presented in USENIX Security 2021} \footnote{https://www.usenix.org/conference/usenixsecurity21/presentation/rakin} 

\end{abstract}

\vspace{-1em}
\section{Introduction}
\vspace{-1em}
Machine Learning (ML), especially deep neural networks (DNN), services in high-performance cloud computing are gaining extreme popularity due to their remarkable performance in intelligent image/video recognition~\cite{lecun1995convolutional,deng2009imagenet,krizhevsky2012imagenet,he2016deep}, natural language processing~\cite{hinton2012deep,lecun2015deep,xiong2016achieving}, medical diagnostics~\cite{deep-ehr}, malware detection~\cite{Droid-Sec}, and autonomous driving~\cite{chen2015deepdriving,MultiNet}. 
Similar to many other high-performance computing (HPC) platforms (e.g., CPU, GPU, ASIC), reconfigurable computing devices like field-programmable gate arrays (FPGA) have been widely deployed in HPC system for DNN acceleration due to their low-effort hardware-level re-programmability to adapt various DNN structures, as well as fast algorithm evolution. For example, IBM and Intel integrated FPGAs in their CPU products for acceleration purposes \cite{IBM_power_sensor_FPGA, Intel_cpu_FPGA}. Alongside the rapid growth of the cloud computing market and critical developments in DNN hardware acceleration, FPGA has become a significant hardware resource for public lease. Recently, the leading cloud service providers have also started integrating FPGAs into their cloud servers. For example, the Stratix-V FPGA from Intel/Altera has been deployed by the Microsoft Project Catapult for DNN acceleration \cite{Microsoft_cloud_FPGA}. Amazon also released its EC2 F1 instances equipped with programmable hardware (UltraScale+VU9P FPGAs) from Xilinx \cite{Amazon_cloud_FPGA}.

For high efficiency and performance, there have been growing efforts to support multiple independent tenants co-residing/sharing an FPGA chip over time or simultaneously \cite{provelengios2020power,zha2020virtualizing}. The \textit{co-tenancy} of multiple users on the same FPGA chip has created a unique attack surface, where many new vulnerabilities will appear and cause dangerous effects. With many hardware resources being jointly used in the multi-tenant FPGA environment, a malicious tenant can leverage such \textit{indirect} interaction with other tenants to implement various new attacks. However, as a relatively new computing infrastructure, as well as one of the main hardware accelerator platforms, the \textit{security of multi-tenant FPGAs for DNN acceleration} has not been investigated in-depth.

From DNN algorithm point of view, its security has been under severe scrutiny through generating malicious input noise popularly known as \textit{Adversarial Examples} \cite{goodfellow2014explaining,madry2017towards,szegedy2013intriguing}. Even though tremendous progress has been made in protecting DNN against adversarial examples~\cite{madry2018towards,carlini2017towards,athalye2018obfuscated}, \textit{neglecting fault injection-based model parameter perturbation does not guarantee the overall security of DNN acceleration in FPGA (DNN-FPGA) system}. Several prior works have effectively demonstrated depletion of DNN intelligence by tempering model parameters (i.e, weights,biases) using supply chain access \cite{Trojannn,gu2017badnets} or through popular memory fault injection techniques \cite{hong2019terminal,Rakin_2019_ICCV,yao2020deephammer,liu2017fault}, which could be in general classified as \textit{adversarial weight attack}. 
Adversarial weight attack can drastically disrupt the inference behavior towards the intent of a malicious party \cite{Rakin_2019_ICCV,yao2020deephammer,rakin2020t,liu2017fault,hong2019terminal}.
The large DNN model's parameters (e.g., weights) are extensively tuned in the training process to play a key role in inference accuracy. 
However, almost all the existing adversarial weight attacks assume an extremely relaxed threat model (i.e., white-box), where the adversary can access all DNN model parameters, like architecture and gradients. Even though it is pivotal to study white-box attacks to understand the behavior of DNN models in the presence of input or weight noise, it is also important to explore how to conduct adversarial weight attacks in a much more strict black-box setup, where the attacker does not know DNN model information.

In summary, \textbf{three primary challenges} are \textbf{i)} Considering multiple tenants \textit{co-reside} on an FPGA, can a malicious user leverage a novel attack surface to provide the luxury of perturbing DNN model parameters of the victim tenant? \textbf{ii)} Can the adversary conduct a black-box adversarial weight attack with no knowledge of DNN model parameters, gradient, etc., instead of white-box attack used in prior works \cite{hong2019terminal,yao2020deephammer}? \textbf{iii)} Given an FPGA hardware fault injection attack scheme and a strict black-box threat model, can an adversary design an efficient searching algorithm to identify critical parameters for achieving a specific malicious objective? Inspired by those challenges, we propose \textit{\mlduplication} attack framework in multi-tenant DNN-FPGA, which consists of \textbf{two main modules}: \textbf{I)} a novel FPGA hardware fault injection scheme, called \textit{adversarial weight duplication (AWD)}, leveraging two different power-plundering circuits to intentionally inject faults into DNN weight packages during data transmission between off-chip memory and on-chip buffer; \textbf{II)} a generic searching algorithm, called \textit{Progressive Differential Evolution Search (P-DES)}, to identify the most vulnerable DNN weight package index and guide AWD to attack for given malicious objective. As far as we know, \mlduplication is the first work demonstrating that the adversarial FPGA tenant could conduct both un-targeted accuracy degradation attack and targeted attack to hijack DNN function in the victim tenant, under both deep learning white-box and black-box setup.
The key contributions of this work are summarized as follows: 

1): The proposed Adversarial weight duplication (AWD) attack is an FPGA hardware-based fault injection method, leveraging the co-tenancy of different FPGA users, to aggressively overload the shared power distribution system (PDS) and duplicate certain DNN model weight parameters during data transmission between off-chip memory and on-chip buffer. Two different power plundering circuits, i.e., Ring Oscillator (RO) and RO with latch (LRO) are explored and validated in the FPGA attack prototype system. 
    
2): To maximize attack efficiency, i.e. conducting AWD-based fault injection into the most vulnerable DNN weight data packages for any given malicious objective, we propose a generic vulnerable weight package searching algorithm, called \emph{Progressive Differential Evolution Search (P-DES)}. It is, for the \emph{first time}, adaptive to both deep learning white-box and black-box setup. Unlike prior works only demonstrated in a deep learning white-box setup \cite{yao2020deephammer}, our success in both white-box and black-box mainly comes from the fact that our proposed P-DES does not require any gradient information of DNN model.
    
3):  We are the first to develop an end-to-end Deep-Dup attack framework, one type of adversarial DNN model fault injection attack, utilizing our DNN vulnerable parameter searching software (i.e. P-DES) to guide and search when/where to inject fault through multi-tenant FPGA hardware fault injection (i.e. AWD) for efficient and effective un-targeted/targeted attacks (i.e., un-targeted attack to degrade overall accuracy and targeted attack to degrade only targeted group accuracy).
    
4): A multi-tenant FPGA prototype is developed to validate the proposed \mlduplication for two different deep learning applications (i.e., Object Detection and Image Classification). Successful un-targeted and targeted attacks are validated and demonstrated in six different popular DNN architectures (e.g. YOLOv2, ResNet-50, MobileNetV2, etc.) on three data sets (e.g., COCO, CIFAR-10, and ImageNet), under both white-box and black-box setups(i.e. attacker has no knowledge of model parameters (e.g. weights/gradients/ architecture)).
    
5): As proof-of-concept, our \mlduplication black-box attack successfully targets the '\emph{Ostrich}' class images (i.e., 100 \% attack success rate) on ImageNet with only \emph{20} (out of \emph{23 Million}) weight package fault injection through AWD attacks on ResNet-50 running in FPGA. Besides, \mlduplication requires just one AWD attack to completely deplete the intelligence of compact MobileNetV2.

\vspace{-1.5em}
\section{Background}
\vspace{-0.5em}
\subsection{Related Attacks on Multi-tenant FPGA}
\vspace{-0.25em}
The re-programmability of FPGA makes it a popular hardware accelerator for customized computing \cite{cong2018customizable}. To further explore the advantages of FPGA, leading hardware vendors like Intel and Xilinx have integrated FPGAs with CPUs \cite{Intel_cpu_FPGA} or ARM cores to build flexible System-on-Chips (SoCs) \cite{Xilinx_Soc,Intel_Soc}. These heterogeneous computing platforms have recently been integrated into cloud data centers \cite{zhao2018fpga}, where the hardware resources are leased to different users. The \textit{co-tenancy} of multiple users on the same FPGA chip, although improves the resource utilization efficiency and performance, but also creates a unique attack surface, where many new vulnerabilities will appear and cause dangerous results. With many critical hardware components (e.g., power supply system) being jointly used in the multi-tenant FPGA environment, a malicious tenant can leverage such \textit{indirect} interaction with other tenants to implement various new attacks. 

Generally, the attacks on multi-tenant FPGAs can be classified into two classes: 1) side-channel attack, in which the adversarial FPGA user can construct hardware primitive as sensors(e.g., ring oscillator (RO)), to track and analyze the secret of victim users. For example, in \cite{zhao2018fpga}, the RO-based sensor used as power side-channel has successfully extracted the key of RSA crypto module, similarly, key extraction from advanced encryption standard (AES) is successfully demonstrated in \cite{krautter2018fpgahammer} based on RO-caused voltage drop. More recently, it has been demonstrated that a malicious user can 
leverage the crosstalk between FPGA long-wires as a remote side-channel to steal secret information \cite{giechaskiel2018leaky,luo2019hill}. 2) Fault injection attack, in which the adversary targets to inject faults to or crash the applications of victim users. For example, the entropy of true random number generator is corrupted by power attacks in multi-tenant FPGAs \cite{mahmoud2019timing}. In \cite{provelengios2019characterization}, the aggressive power consumption by malicious users causes a voltage drop on the FPGA, which can be leveraged to introduce faults. 

With Machine Learning as a service (MLaaS) \cite{AWS_MLaaS,Cloud_AutoML} becoming popular, public lease FPGAs also become an emerging platform for acceleration purposes. However, the security of using multi-tenant FPGA for DNN acceleration is still under-explored in existing works, which is the main target of this paper. Specially, the proposed \mlduplication methodology belongs to the fault injection category, which leverages malicious power-plundering circuits to compromise the integrity of the DNN model for un-targeted or targeted attacks.

\vspace{-0.75em}
\subsection{Deep Learning Security}
\vspace{-0.25em}
There has been a considerable amount of effort in developing robust and secure DL algorithms~\cite{goodfellow2014explaining,madry2017towards,carlini2017towards,gu2017badnets,abs-1708-08689,ilyas2018blackbox,cubuk2017intriguing,Ateniese15,Tramer16,Papernot17,Fredrikson15,fredrikson2014privacy}. Existing deep learning attack vectors under investigation mainly fall into three categories: \emph{1)} Attacks that either mislead prediction outcome using maliciously crafted queries (i.e., adversarial inputs/examples~\cite{narodytska2016simple,carlini2017towards}) or through miss-training the model with poisoned training set (i.e., data poisoning attacks~\cite{biggio2013poisoning,xiao2015feature}).
\emph{2)} DL information leakage threats such as membership inference attacks~\cite{Shori17,fredrikson2014privacy} and model extraction attacks~\cite{Papernot17,kesarwani2018model} where adversaries manage to either recover data samples used in training or infer critical DL model parameters. \emph{3)} Finally, adversarial fault injection techniques have been leveraged to intentionally trigger weight noise to cause classification errors in a wide range of DL evaluation platform \cite{hong2019terminal,liu2017fault,Rakin_2019_ICCV,yao2020deephammer,breier2018deeplaser}.

The first two attacks are generally considered as \emph{external adversaries} that exploit training and inference inputs to the deep learning model. Despite the progress in protecting DNN against this external adversaries~\cite{madry2018towards,carlini2017towards,athalye2018obfuscated}, neglecting internal adversarial fault injection still puts the overall security of DNN acceleration in FPGA (DNN-FPGA) systems under threat. The most recent adversarial weight attacks \cite{Rakin_2019_ICCV,rakin2020t,rakin2020tbt,yao2020deephammer} demonstrated, in both deep learning algorithm and real-word general-purpose computer system, that it is possible to modify an extremely small amount (i.e., tens out of millions) of DNN model parameters using row-hammer based bit-flip attack in computer main memory to severely damage or hijack DNN inference function. Even those injected faults might be minor if leveraged by a malicious adversary, such internal \emph{adversarial fault injection} harnessing hardware vulnerabilities may be extremely dangerous as they can severely jeopardize the confidentiality and integrity of the DNN system.  

\begin{figure}[tb!]
    \centering
    \includegraphics[width=1\linewidth]{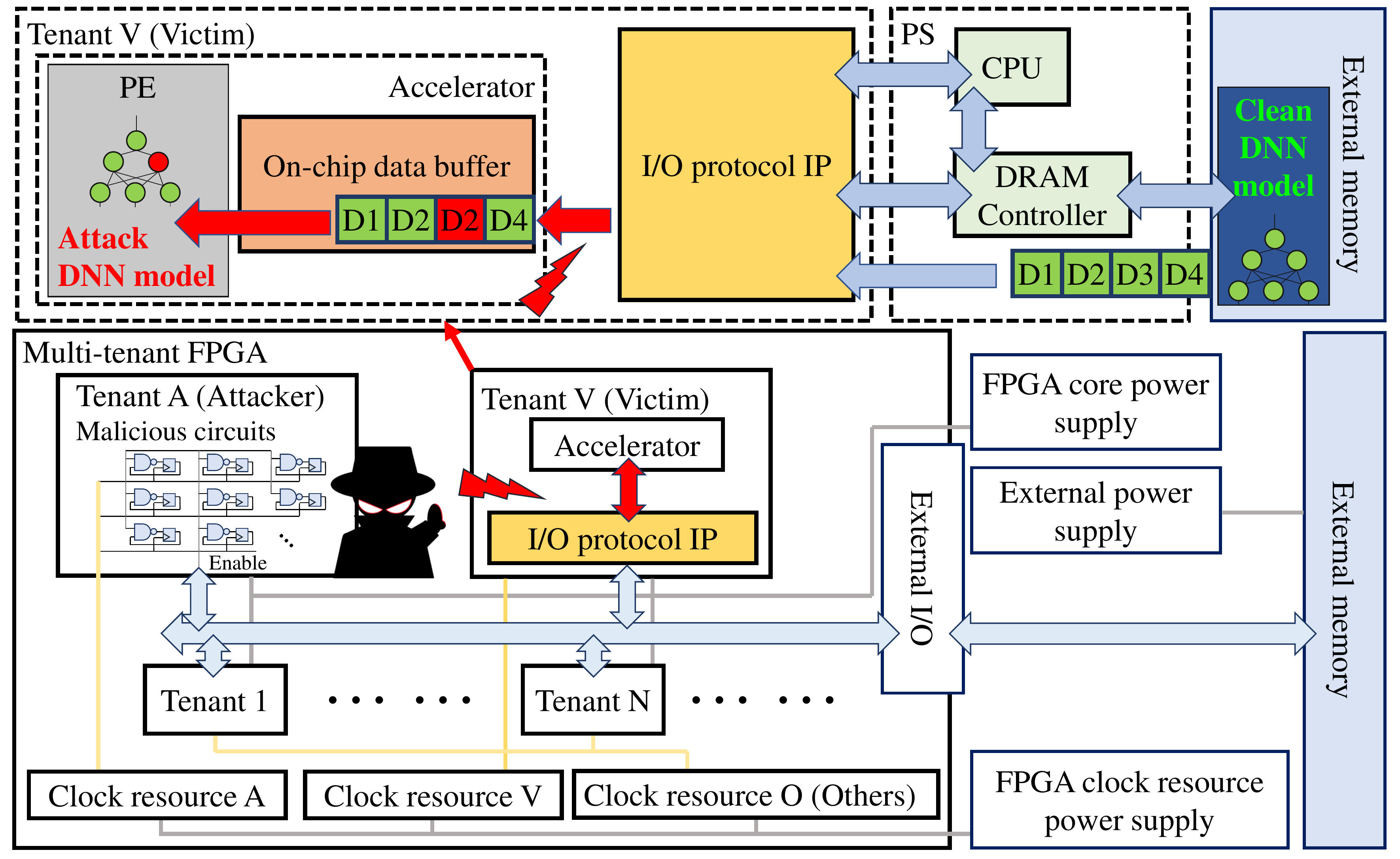} 
    \caption{Threat model for the proposed \mlduplication.
    }
    \label{fig:threat_model}
    \vspace{-1.5em}
\end{figure}

\vspace{-0.5em}
\section{Threat Model and Attack Vector}\label{sec:threat_model}
\vspace{-0.5em}


\textbf{Multi-tenant FPGA Hardware Threat Model.} In this work, we consider the representative hardware abstraction of multi-tenant FPGA used in the {\textit{security}} works \cite{ramesh2018fpga,giechaskiel2018leaky,yazdanshenas2019costs}, and {\textit{operating system}} works \cite{khawaja2018sharing,zha2020virtualizing}. The threat model is shown in Fig. \ref{fig:threat_model}, which has the following characteristics: (1) Multiple tenants \textit{co-reside} on a cloud-FPGA and their circuits can be executed simultaneously. The system administrator of cloud service is trusted. (2) Each tenant has the flexibility to program his design in the desired FPGA regions (if not taken by others). (3) All tenants share certain hardware resources on an FPGA chip, such as the PDS and the communication channels with external memory or I/O. 
(4) We assume that the adversary knows the type of transmitted data (i.e., either DNN model or input data) on the communication channel (e.g., I/O protocol IP) connecting the off-chip memory and on-chip data buffer. Adversarial FPGA tenants can learn such information in different ways: i) Using the side-channel leakage from the communication/data channels on the FPGA, e.g., the cross-talk between FPGA long-wires \cite{giechaskiel2018leaky}. Besides, recent works have reverse engineered DNN using side-channel attacks to practically recover its information (i.e, architecture, weights) \cite{batina2019csi,moini2020remote}. Additionally, it is practical to recover the DNN model using instruction flow leakage \cite{hu2020deepsniffer}. ii) Practically, the victim FPGA tenant can be the provider of Machine learning as a service (MLaaS)\cite{AWS_MLaaS,Cloud_AutoML}, who offer accelerated DNN computation on multi-tenant FPGA, and the adversary can rent such service as a normal customer, then he/she can learn some info of the model and query outputs. More importantly, our black-box attack only requires to know the transmitted data type (i.e. weight or input), instead of actual weight values, which is recoverable using similar methods as in \cite{batina2019csi,moini2020remote,giechaskiel2018leaky}. It is worth mentioning that, although the current cloud-computing business model has not yet supported simultaneous resource-sharing, with the significant development of FPGA-based cloud computing, e.g., dynamic workload support~\cite{khawaja2018sharing}, FPGA virtulization~\cite{knodel2016rc3e}, multi-tenant FPGA is envisioned to be possible in the future \cite{yazdanshenas2019datacenter}.

\textbf{Deep Learning (DL) Algorithm Threat Model.}
Regarding the Deep Learning algorithm level threat model, in this work, following many prior DL security works \cite{goodfellow2014explaining,madry2018towards,he2019parametric,yao2020deephammer,Rakin_2019_ICCV,hong2019terminal,rakin2020tbt,rakin2020robust}, two different DL algorithm threat models are considered and defined here: 1) \textit{DL white-box}: attacker needs to know model architectures, weight values, gradients, several batches of test data, queried outputs. 2) \textit{DL black-box}: attacker only knows the queried outputs and a sample test dataset. Unlike the traditional DL white-box threat model \cite{goodfellow2014explaining,madry2018towards,Rakin_2019_ICCV,buckman2018thermometer}, our DL white-box is even weaker with no requirement of computing gradient during the attacking process. Since different DL security works may have different definitions of white/black-box, throughout this work, we will stick to the definition here, which is commonly used in prior works \cite{Rakin_2019_ICCV,buckman2018thermometer,chen2017zoo}. In this work, similar to many adversarial input or weight attacks, we only target to attack a pre-trained DNN inference model in FPGA, i.e., hijacking the DNN inference behavior through the proposed \mlduplication, not the training process, which typically requires extra access to the training supply chain \cite{Trojannn,badnets}.

In our threat model defined in Fig. \ref{fig:threat_model}, the adversary will leverage our proposed AWD based fault injection attack on the weight packages identified by our proposed P-DES searching algorithm, when transmitting the DNN model from off-chip memory to on-chip buffer/processing engine (PE), resulting in a weight perturbed DNN model in the PEs. After the attack, the DNN function is hijacked by an adversary with malicious behaviors, such as accuracy degradation or wrong classification of a targeted output class.
\vspace{-1.5em}
\section{Attack Objective Formulation}
\vspace{-0.75em}
\label{sec: attackobj}
The proposed \mlduplication attack is designed to perform both un-targeted and targeted attacks, defined as below.

\textbf{Un-targeted Attack.}
The objective of this attack is to degrade the overall network inference accuracy (i.e., miss-classifying whole test dataset), thus maximizing the inference loss of DNN. As a consequence, the objective can be formulated as an optimization problem:
\vspace{-0.5em}
\begin{align}
\max~\mathcal{L}_u = \max_{\{\hat{W}\}}~\E_{\mathbb{X}} \mathcal{L}(f(\vx, \{W\}); \vt) 
\label{eqt:unt}
\vspace{-5em}
\end{align}
where $\vx$ and $\vt$ are the vectorized input and target output of a given test batch and $\mathcal{L}(\cdot,\cdot)$ calculates the loss between DNN output and target. The objective is to degrade the network's overall accuracy as low as possible by perturbing weights of the clean DNN model from $W$ to $\hat{W}$.

\textbf{Targeted Attack.}
Different from the un-targeted attack, the objective of targeted attack in this work is to misclassify a specific (target) class of inputs ($t_s$). This attack objective is formulated in Eq. \ref{eqt:loss_T-BFA_yolo}, which can be achieved by maximizing the loss of those target class: 
\begin{align}
\max~\mathcal{L}_t = \max_{\{\hat{W}\}}~\E_{\mathbb{X}} \mathcal{L}(f(\vx_s, \{W\}); \vt) 
\label{eqt:loss_T-BFA_yolo}
\vspace{-0.75em}
\end{align}
where $\vx_s$ is a sample input batch belongs to the target class $t_s$. 
\vspace{-2em}
\section{Proposed Deep-Dup Framework}
\vspace{-1em} 
\textit{\mlduplication} mainly consists of two proposed modules: 1) \textit{adversarial weight duplication (AWD)} attack, a novel FPGA hardware fault injection scheme leveraging power-plundering circuit to intentionally duplicate certain DNN weight packages during data transmission between off-chip memory and on-chip buffer; 2) \textit{progressive differential evolution search (P-DES)}, a generic searching algorithm to identify most vulnerable DNN weight package index and guide AWD fault injection for given malicious objective. In the end of this section, we will present \mlduplication as an end-to-end software-hardware integrated attack framework. 
\vspace{-1em}
\subsection{AWD attack in multi-tenant FPGA}
\vspace{-0.25em}
\subsubsection{Preliminaries of DNN model implementations}
\vspace{-0.5em}

The schematic of an FPGA-based DNN acceleration is illustrated in Fig. \ref{fig:threat_model}, consisting of a processing system (PS), processing engine (PE), and external (off-chip) memory. Practically, 
DNN computation is usually accomplished in a \textit{layer-by-layer} style, i.e., input data like image and DNN model parameters of different layers are usually loaded and processed separately \cite{zhang2015optimizing,zhang2020dnnexplorer,xu2020autodnnchip}. 
Fig. \ref{fig:threat_model} shows the flow of FPGA I/O protocol IP for typical DNN model transmission, in which the on-chip data buffer sends a data transaction request to PS for loading data from external memory. Then, the processing engine (PE) will implement computation based on the DNN model in the on-chip data buffer (e.g., BRAM). 

\begin{figure}[tb!]
    \centering
     \begin{subfigure}{0.54\linewidth}
        \centering
        \includegraphics[width=1\linewidth]{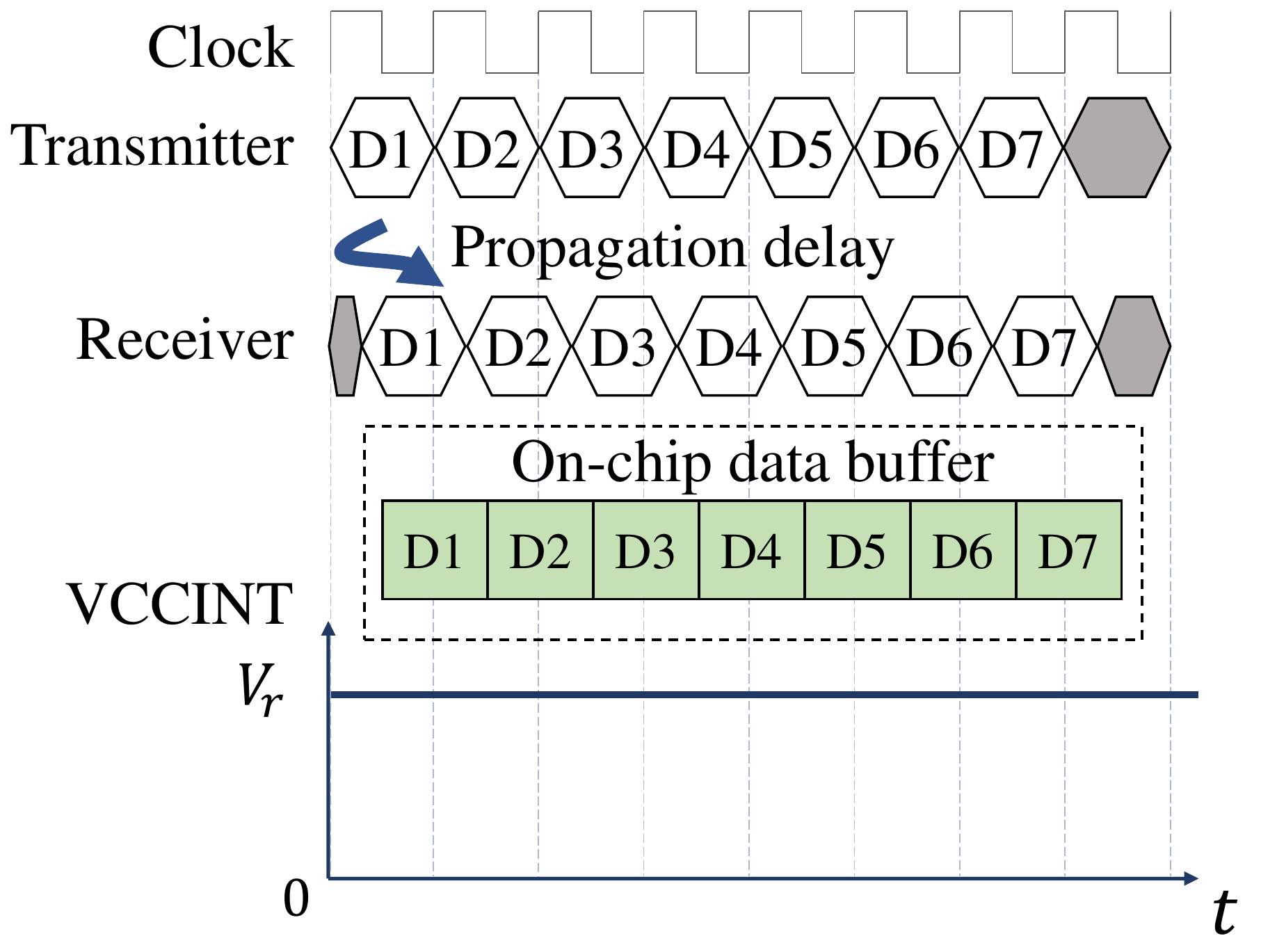}
        \subcaption{DNN model transmission w/o attack.}
        \label{Fig:no_atk}
     \end{subfigure}
     \hspace{1pt}
     \begin{subfigure}{0.43\linewidth}
        \centering
        \includegraphics[width=1\linewidth]{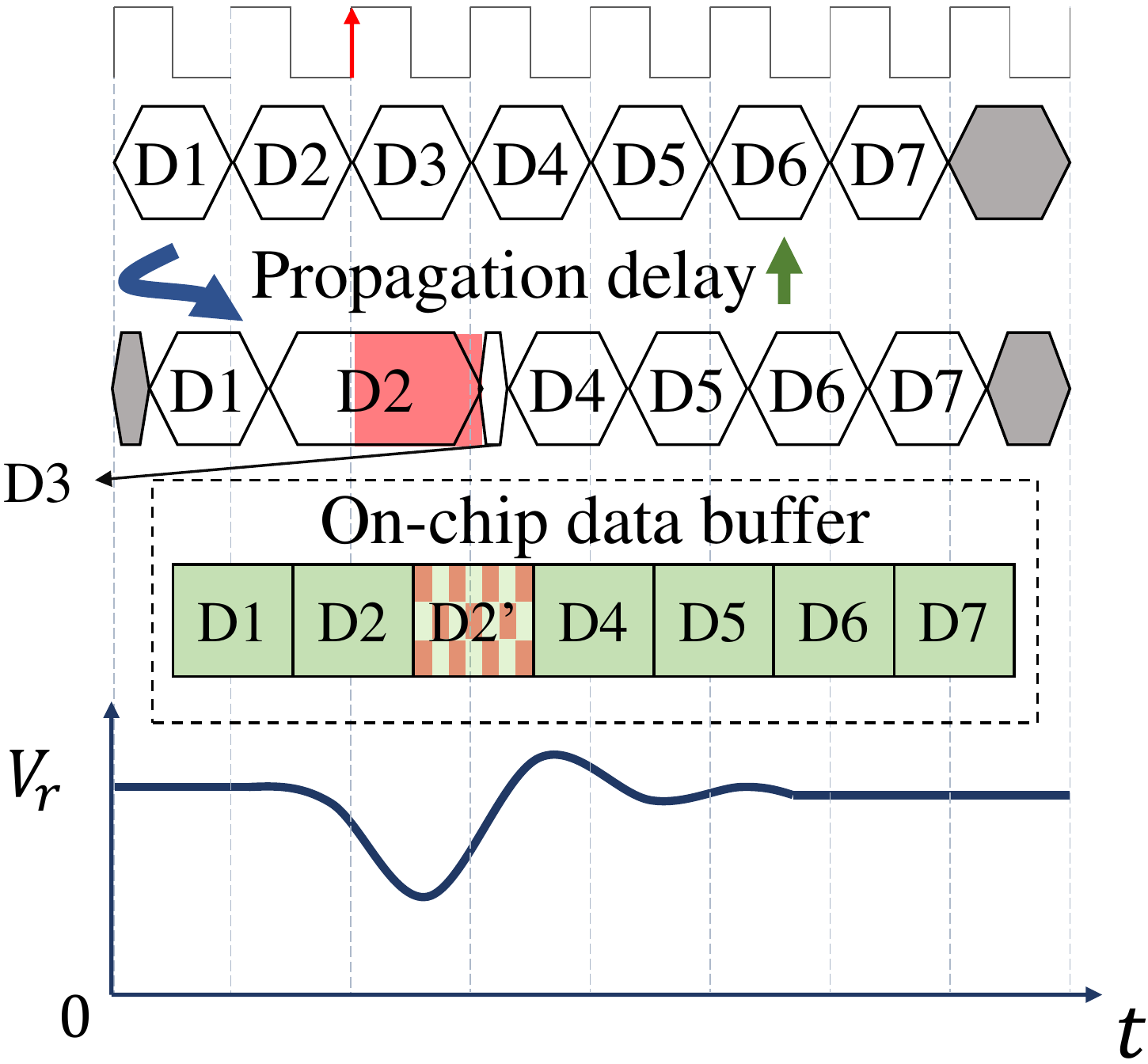}
        \subcaption{DNN model transmission under AWD attack.}
        \label{Fig:weak_atk}
    \end{subfigure}
    \caption{Illustrated timing diagrams of DNN model transmission w/o or under AWD attack. (a) Each DNN weight package (D$i$) is transmitted and received in a separate clock cycle. (b) Voltage glitch incurs more propagation delay to the transmission of D2, which also shortens the next package D3. As a result, the data package D2 is sampled twice by the receiver clock, injecting faults to the received data package.} 

    \label{Fig:On-chip_buffer_status}
    \vspace{-2em}
\end{figure}

A data transmission flow is shown in Fig. \ref{Fig:no_atk}, in each clock cycle, a \textit{data package} (D$i$) is transmitted from transmitter (e.g. external memory) to receiver. 
Taking the advanced eXtensible interface4 (AXI4) as an example \cite{AMBA_AXI}, 
the receiver first sends a data request with an external memory address, and then it will be notified to read the data when it is ready. The size of each transmitted data package depends on the channel bandwidth. In DNN model transmission, the normal (w/o attacks) transmission flow with each D$i$ as a DNN weight package is illustrated in Fig. \ref{Fig:no_atk}, with FPGA core voltage (\texttt{VCCINT}) being stable at the recommended supply voltage ($V_r$), $N$ data packages (e.g., weights) are transmitted in $N$ clock cycles (D1-D7 in Fig. \ref{Fig:no_atk}).

\vspace{-1em}
\subsubsection{AWD based fault injection into DNN model} 
\vspace{-0.5em}
The power supply of modern FPGA chips is regulated based on their voltages, different components will be activated following the order of their nominal voltage, e.g., from low to high \cite{xilinx2018_DS181,xilinx2019_DS183,xilinx2019_DS925}. Most FPGAs utilize a hierarchical power distribution system (PDS) \footnote{PDS is the official terminology of Xilinx FPGAs, while Intel FPGAs use power distribution networks \cite{Intel_PDN}. For uniformity, we use PDS in this paper.}, which consists of some power regulators providing different supply voltages \cite{TPS54620,xilinx2019_DS183,xilinx2019_DS925}. A critical component of PDS is the capacitor used as the ``power bank'' for the operational reliability of FPGA. For example, when an FPGA chip's power supply is suddenly overloaded (i.e., by a transient higher power demand), these capacitors are discharged to compensate for the extra power that regulators cannot immediately provide. The capacitors of FPGA PDS are usually sized accordingly to fit the practical need. Formally, the default output capacitance ($C_{out}$) of an FPGA is usually sized to compensate for the current difference for at least two clock cycles with a tolerable voltage drop \cite{TPS54620}. As calculated in Eq.~\ref{eq:capacitance}, where $\Delta I_{out}$ and $\Delta V_{out}$ represent the changes of output current and voltage, respectively, and $f_{sw}$ denotes the regulator switching frequency. 
\vspace{-0.5em}
\begin{equation}\label{eq:capacitance}
C_{out} = \frac{2\times \Delta I_{out}}{f_{sw}\times \Delta V_{out}}
\vspace{-0.5em}
\end{equation}
As one of FPGA's most critical parameters, the clock signals provide standard and global timing references for all on-chip operations. In practice, to generate different timing signals, i.e., with different frequencies or phases, FPGAs are equipped with several clock management components, such as the phase-lock-loop. The on-chip clock signals are usually generated by various clock management components, and their reliability is heavily dependent on the robustness of these components. To enhance clock integrity, these clock components are powered by separate supply voltage resources (Fig. \ref{fig:threat_model}) from the computing elements like PE. For example, the clock components of Xilinx FPGAs are powered by the auxiliary voltage \texttt{VCCAUX} rather than the FPGA core supply voltage \texttt{VCCINT} \cite{xilinx2020_UG583}. Such a separate power supply mechanism ensures sufficient energy for the operation of these clock components, thus enhancing reliability. 

The DNN execution in FPGA is significantly relying on the integrity of its loaded model. Our proposed AWD attack is motivated by two facts: 1) As aforementioned, the reliability and correctness of FPGA applications are ensured by the power delivery mechanism; 2) Based on the power regulation mechanism, there exists a \textit{{maximum power capacity}} that FPGA PDS can provide to PEs. Thus, if the FPGA PDS is overloaded, FPGA applications might encounter faults caused by the timing violation between the clock signal and computation/data. Recent works have demonstrated that the activation of many power-plundering circuits (e.g., ROs), can cause transient voltage drop on the FPGA \cite{krautter2018fpgahammer,gnad2017voltage,mahmoud2019timing}, thus incurring fault injection. 

Considering the importance of frequent and real-time DNN model transmission from/to FPGA, the basic idea for AWD attack is that a malicious FPGA tenant can introduce a timing violation to the DNN model transmission from off-chip memory to the on-chip data buffer. As illustrated in Fig. \ref{Fig:no_atk}, a stable FPGA core voltage (\texttt{VCCINT}) (i.e., with trivial or no fluctuations) will not cause timing violations to data transmission. However, an unstable \texttt{VCCINT} will incur serious timing violations. For example, a sudden voltage drop will make the digital circuit execution slower than usual, causing a longer propagation delay to the data transmission. As shown in Fig. \ref{Fig:weak_atk}, the adversary's aggressive power plundering creates a voltage drop/glitch that incurs slowing down the data transmission channel. As a result, the corresponding data package (e.g., D$2$) may be sampled twice by the receiver clock, causing a fault injection into the following data package. We envision that maliciously designed fault-injected weight data packages will greatly impact the DNN computation, inducing either significant performance loss, or other malicious behaviors.

\vspace{-1em}
\subsubsection{Power-plundering circuits}
\vspace{-0.6em}
\begin{figure}[tb!]
    \centering
    \begin{subfigure}{0.35\linewidth}
       \centering
        \includegraphics[width=\textwidth]{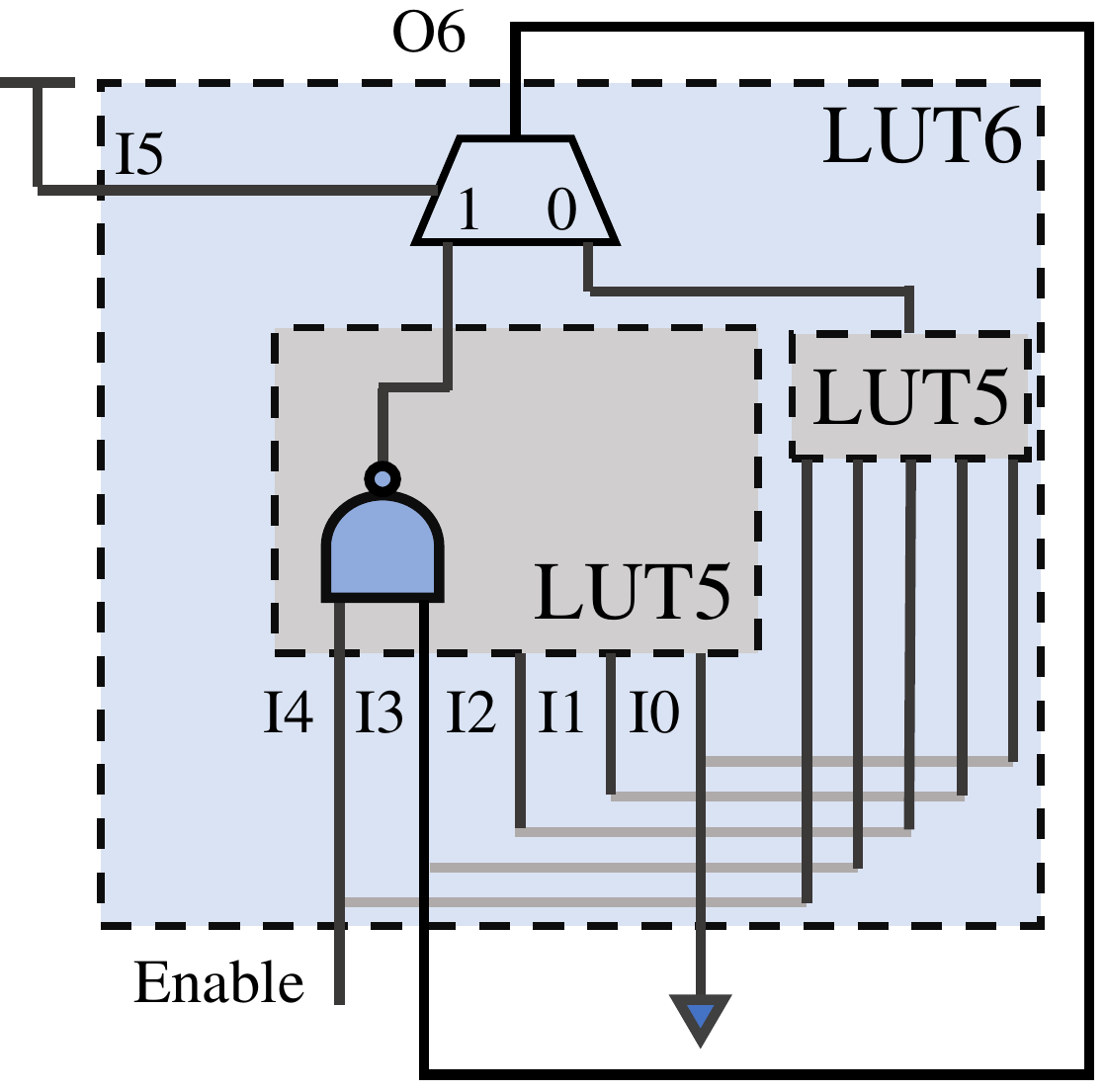}
        \subcaption{A power-plundering cell based on ring-oscillator (RO).}
        \label{Fig:PW-RO}
    \end{subfigure}
    \hspace{10pt}
    \begin{subfigure}{0.5\linewidth}
        \centering
        \includegraphics[width=\linewidth]{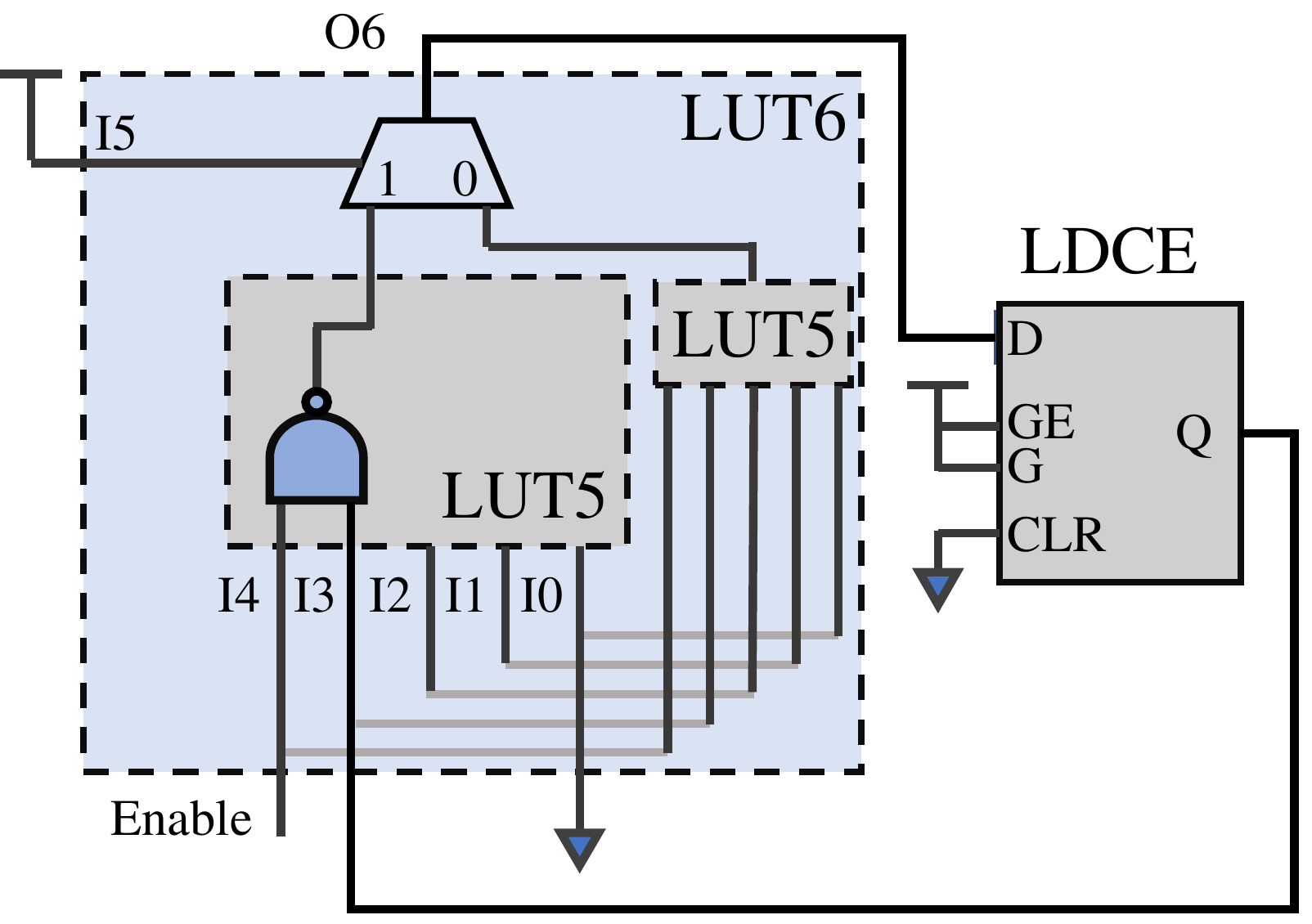}
        \subcaption{A cloud-sanctioned power-plundering cell based on RO with a Latch (LRO).}
        \label{Fig:PW-LRO}
    \end{subfigure}
    \vspace{-1em}
    \caption{Two power-plundering circuit examples on FPGA}. 
    
    \label{Fig:PW_circuit}
    \vspace{-1.5em}
\end{figure}
A power-plundering circuit can be achieved with any circuit scheme with high dynamic power consumption, e.g., ring-oscillator (RO) circuits. However, it should be noted that although RO circuit provides high power-plundering potential, it can be possibly detected by the FPGA development tools \cite{la2020fpgadefender}. To make power-plundering more stealthy, i.e., cloud-sanctioned, some recent works employ common FPGA applications, e.g., the shift registers of an AES circuit \cite{provelengios2020power} and XOR tree circuit \cite{matas2020power}. Since this work focuses on the security of the DNN model in multi-tenant FPGA, we adopt two power-plundering schemes, RO and Latch RO (LRO), for proof-of-concept. Fig. \ref{Fig:PW-RO} shows the RO circuit instantiated with an FPGA look-up table (LUT). Different from RO, the LRO circuit shown in Fig. \ref{Fig:PW-LRO} has a latch in the loop, which is a cloud-sanctioned design scheme that can bypass the design rule checking for combinational loop in FPGA design tools. 
In detail, these two power-plundering schemes are both instantiated as a NAND gate controlled by an \texttt{Enable} signal. 
An adversarial FPGA tenant can employ a large number of such cells controlled by the same \texttt{Enable} signal, which can be activated to overload the FPGA PDS and introduce transient voltage drop shown in Fig. \ref{Fig:weak_atk}, thus implementing fault injection attack. Note that the proposed attack in this paper can be achieved with any other cloud-sanctioned power plundering design, such as the AES-based scheme in \cite{provelengios2020power}.


\vspace{-1em}
\subsubsection{{AWD attack triggering system}}\label{sec:triggering_sys}
\vspace{-0.6em}

\begin{figure}[tb!]
    \centering
    \includegraphics[width=0.9\linewidth]{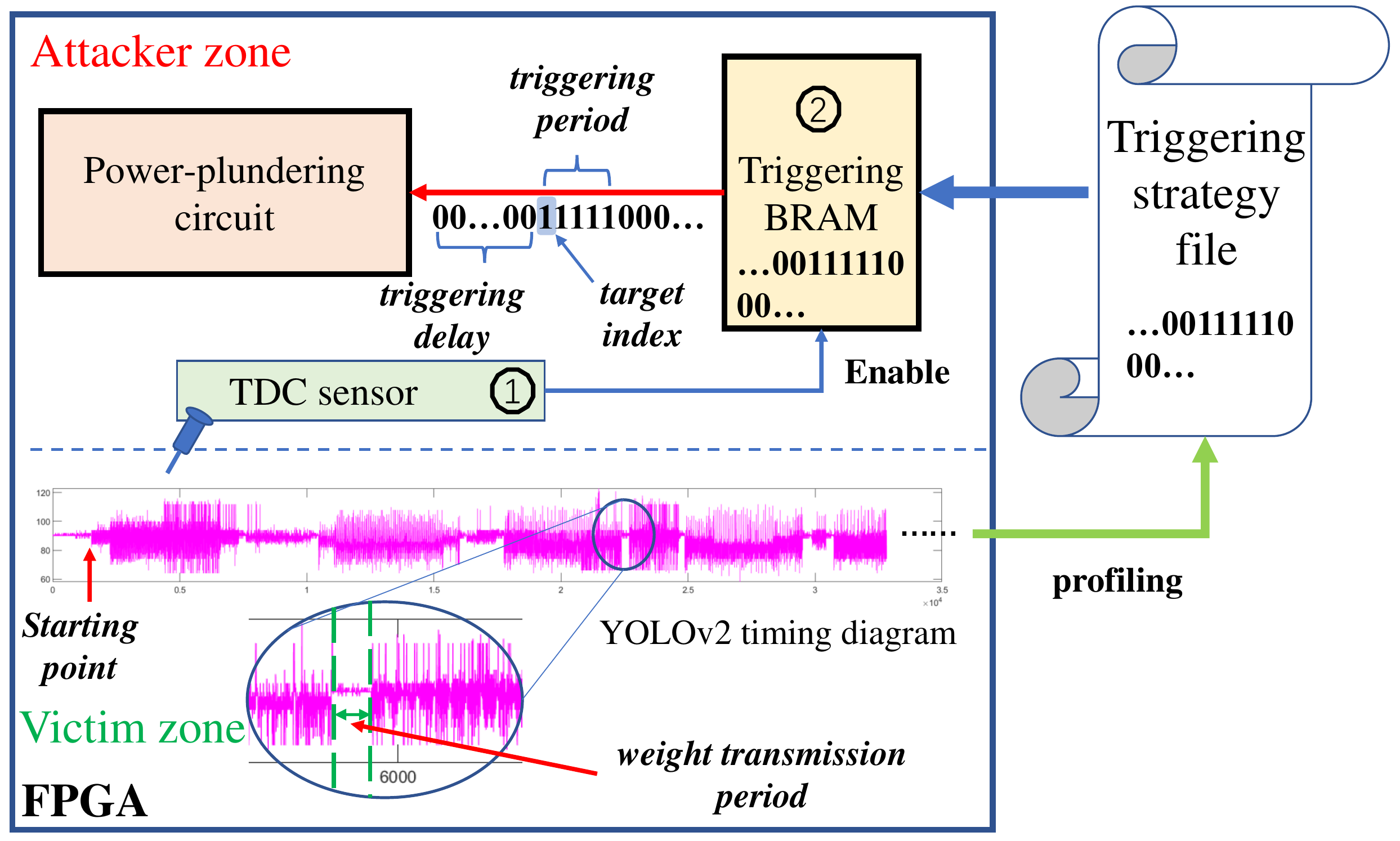} 
    \caption{AWD triggering system. A TDC sensor is used to capture voltage fluctuation during the YOLOv2 execution, in which the weight transmission period can be clearly observed.} 

    \label{fig:trigger_img}
    \vspace{-1.5em}
\end{figure}

As mentioned in the hardware threat model (Sec.\ref{sec:threat_model}), our proposed attack only requires the adversary to know the type of data (i.e., weight or not) being transmitted on the FPGA and the \textbf{starting/ending points}, which can be achieved with side-channel (e.g., power) analysis. 
To demonstrate this, we build the AWD triggering system with two major components: \circled{1} \textit{Time-to-Digital Converter} (TDC) based sensor and \circled{2} \textit{Triggering BRAM}, as shown in Fig. \ref{fig:trigger_img}. We prototype a TDC circuit in FPGA to capture the on-chip voltage fluctuation and measure the digital output of the TDC sensor during the execution of DNN (YOLOv2 in this example). We observe a strong correlation between the sensor outputs and DNN execution, i.e., weight transmission or functional layers' execution. For example, as shown in Fig. \ref{fig:trigger_img}, the TDC sensor outputs corresponding to weight transmission periods are relatively stable (i.e., much less voltage fluctuation), since it consumes much less power than the functional layers, like Max pool or Convolution. Due to the page limit, we omit the TDC sensor design details and refer interested readers to the related work  \cite{luo2021deepstrike} for details.

Based on the TDC sensor output, we profile a \textit{triggering strategy file} to control the AWD attack activation, which consists of three parameters: triggering delay, triggering period, and target index. The strategy file is stored in the triggering BRAM (\circled{2}), composed of `1s' and `0s, which are used to activate or disable the power-plundering circuit, respectively. With the triggering BRAM being read at a certain clock frequency, this system can control the triggering of fault injection. For example, a series of consecutive `0s' disable the power plundering circuit for a certain time period, while a series of consecutive `1s' defines the length of the attack period. By selecting the locations of `1s', we can choose to inject faults on specific DNN weights of specific attack indexes obtained from our P-DES searching algorithm (Sec.\ref{sec:p_des}). 

\vspace{-0.75em}
\subsection{P-DES Searching Algorithm} \label{sec:p_des}
\vspace{-0.25em}
This section delineates the proposed vulnerable weight searching algorithm, called \textit{Progressive Differential Evolution Search (P-DES)}, to generate a set of weight data package index for AWD to attack, given attack objective. To formally define the problem, let us first consider a $L$ layer network with weight parameters-${W}_{l=1}^L$. Then, the after-attack (i.e. perturbed) weight of the target DNN model executed in FPGA will become $\hat{{W}}_{l=1}^L$. We model different attack objectives aiming to minimize the difference between ${W}_{l=1}^L$ and $\hat{{W}}_{l=1}^L$ for deriving the minimal number of required AWD attacks performing both defined un-targeted and targeted attack objectives.


To clearly describe the searching algorithm, we start from modeling of white-box attack, assuming attacker knows the exact model parameters (i.e. weight values and architecture). The black-box attack will leverage a similar searching algorithm and its corresponding adaption will be described in the end-to-end attack framework section. We assign each weight package in the target DNN with two indexes $(p,q)$; where $p$ denotes the layer index and $q$ denotes the index of weight at layer $p$ after flattening the weight matrix $\textbf{W}$ ($\textbf{W} \in R^{m\times n \times a \times kw}$) into a 1D array. Note that, here the weight package refers to one data package that is transmitted in one clock cycle. 
In the following, we may just call it weight for simplification. 
The proposed search algorithm is general and applicable for both attack objectives described in Sec. \ref{sec: attackobj}. 

\begin{figure}[tb!]
    \centering
    \includegraphics[width=0.4\textwidth]{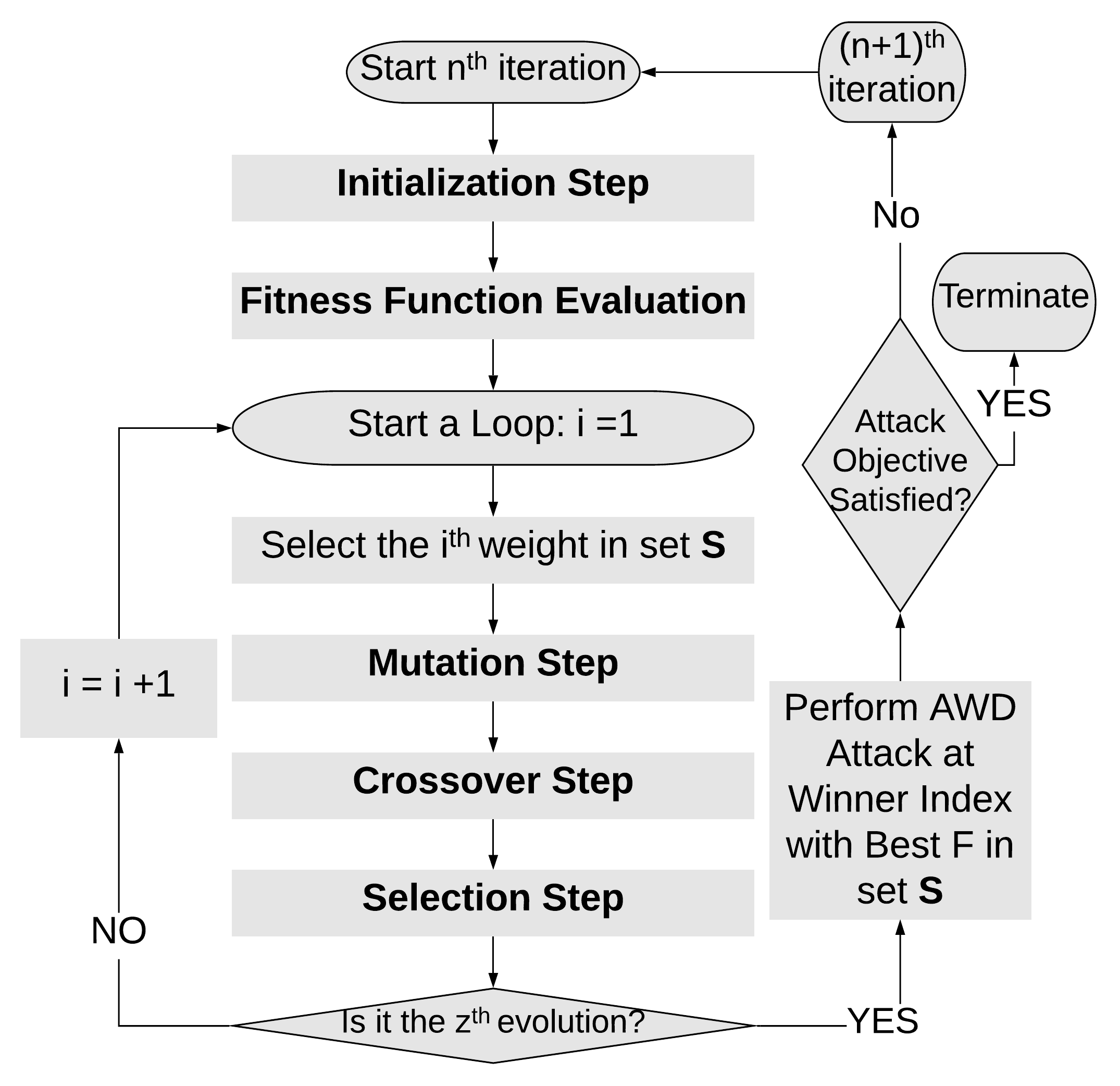} 
    \caption{Overview of  proposed adversarial weight index searching (P-DES) algorithm.}
    \label{fig:DES}
    \vspace{-1.5em}
\end{figure}


P-DES is a progressive search algorithm integrating with the concept of differential evolution \cite{mayer2005differential,price2013differential,zhang2011improved}. The goal is to progressively search for one weight index at each iteration to guide AWD attack until the attacker-defined malicious objective is satisfied. The flow chart of the proposed P-DES is shown in Fig. \ref{fig:DES}. For $n^{th} iteration$, it starts by initializing a set of random weight candidates (i.e. population set - $\textbf{S}$) for attacker to perform AWD attack and evaluate each attack effect (i.e. fitness function) at current iteration. Then it runs through a succession of evolutionary steps: \emph{mutation}, \emph{crossover} and \emph{selection} for $z$ times (known as the number of \emph{evolution}, '500' in this work) to gradually replace original candidates with better ones for achieving the attacker defined malicious objective. 
When $z$ times evolution is finished in one search iteration, the attacker picks one best candidate (weight index with highest fitness function value- $F$) among the final survived population set $S$ and conduct an AWD attack on this winner weight location to duplication data package as described in the previous sub-section. The detailed description of each step is as follow:

\vspace{-1em}
\paragraph{Initialization Step.}
As described above, the objective of differential evolution is to improve population set $\textbf{S}$ over time to gradually reach the attacker-defined malicious objective. To initialize, $\textbf{S}$ will start with a set of random values, containing $z$ weights whose indexes located at $(p_l,q_l)$ ; where $l=1,2,3,..,z$. Here, $z$ is the size of $\textbf{S}$, defined as the number of evolution. Ideally, a larger population set (i.e., higher $z$) would result in a better attack performance at the cost of increased searching time.
\vspace{-1.5em}
\paragraph{Fitness Function Evaluation.} Fitness function - $F_l$ is an important step of an evolutionary algorithm to evaluate the attack effect of each proposed candidate in the population set $\textbf{S}$. In our \mlduplication attack, as defined in Eq. \ref{eqt:unt} and Eq. \ref{eqt:loss_T-BFA_yolo}, we assign the DNN loss function as fitness function. Thus we could evaluate the attack effect (i.e. $F_l$) of each candidate in set $\textbf{S}$ in terms of DNN loss. Note that, for a white-box attack, such evaluation (i.e. fitness function) could be computed in an off-line replicated model. For black-box attack, the loss will be directly evaluated in FPGA by conducting AWD attack in the proposed candidate index pointed data package clock. In the next sub-section, a detailed \mlduplication framework for both white-box and black-box attacks will be discussed. In P-DES, the attacker's goal is to maximize the fitness function - $F_l$ to achieve un-targeted (Eq. \ref{eqt:unt}) or targeted attack (\ref{eqt:loss_T-BFA_yolo}):

\begin{equation}
    F_l \in \{\mathcal{L}_u,\mathcal{L}_t \}
    \label{Eqn: fit}
    \vspace{-0.25em}
\end{equation}
where ${L}_u$ is un-taregeted attack loss and ${L}_t$ is targeted attack loss.
Note that, 
the after each evaluation of $F_l$, attacker needs to restore the original weight values $W$ by reloading the weights, to guarantee each fitness function is evaluated only based on one corresponding attack weigh index.

\paragraph{Mutation Step.}
For each weight index candidate in population set $S$, the mutation step generates new candidates using specific mutation strategy to improve current population set. In this work, we integrate four popular mutation strategies\cite{das2016recent,vitaliy2006differential}, where each one generates one mutant vector. Thus, a mutant vector ( \{$p_{mut},q_{mut}$\} =\{($p_{mut1},q_{mut1}$);($p_{mut2},q_{mut2}$);($p_{mut3},q_{mut3}$);($p_{mut4},q_{mut4}$\} )) is generated for each weight index candidate:

\noindent\textbf{Strategy 1:}
\vspace{-0.5em}
\begin{align}
    p_{mut1} = p_{a} + \alpha_1 (p_b-p_c);  \\
    q_{mut1} = q_{a} + \alpha_1  (q_b-q_c) 
    \label{Eqn:1}
    \vspace{-0.25em}
\end{align}
\vspace{-0.25em}
\noindent\textbf{Strategy 2:}
\vspace{-0.25em}
\begin{align}
    p_{mut2}=p_{a}+ \alpha_1 \times (p_b-p_c) + \alpha_2 \times (p_d-p_e); \\
    q_{mut2}=q_{a}+ \alpha_1 \times (q_b-q_c) +  \alpha_2 \times (q_d-q_e) 
    \vspace{-0.25em}
\end{align}
\vspace{-0.25em}
\noindent\textbf{Strategy 3:}
\begin{align}
    p_{mut3} = p_{a}+ \alpha_1  (p_{best}-p_a) + \alpha_2 (p_b-p_c) +  \alpha_3 (p_d-p_e);  \\
    q_{mut3} = q_{a}+ \alpha_1(q_{best}-q_a) +  \alpha_2(q_b-q_c)  +  \alpha_3(q_d-q_e) 
    \vspace{-0.25em}
\end{align}
\vspace{-0.5em}
\noindent\textbf{Strategy 4:}
\vspace{-0.25em}
\begin{align}
    p_{mut4} = p_{a} + \alpha_1 (p_{best}-p_{worst}); \\
    q_{mut4} = q_{a} + \alpha_1  (q_{best}-q_{worst}) 
    \label{Eqn:2}
    \vspace{-0.25em}
\end{align}
where $\alpha_1,\alpha_2,\alpha_3$ are the mutation factors sampled randomly in the range of [0,1] \cite{das2016recent}. $a,b,c,d,e$ are random numbers ($a \neq b \neq c \neq d \neq e$) generated in the range of [0,z]. ($p_{best},q_{best}$) and ($p_{worst},q_{worst}$) are the indexes with the best and worst fitness function values. Note that, both $p$ and $q$ for each layer are normalized to the range of [0,1], which is important since the amount of weights at each layer is different.

\vspace{-0.25em}
\paragraph{Crossover Step.} 
In the crossover step, attacker mixes each mutant vector ($p_{mut},q_{mut}$) with current vector ($p_i,q_i$) to generate a trial vector($p_{trail},q_{trial}$): 
\vspace{-0.25em}
\begin{align}
    if \ p_{mut} \in [0,1]: \ p_{trial} = p_{mut};\quad else: \
    p_{trial} = p_i \\
    if \ q_{mut} \in [0,1]: \ q_{trial} = q_{mut};\quad else: \
    q_{trial} = q_i 
    \vspace{-0.25em}
\end{align}
 
The above procedure guarantees attacker only chooses the mutant feature with a valid range of [0,1]. 
Then, the fitness function is evaluated for each trial vector (i.e., $F_{trial1}$,$F_{trial2}$,$F_{trial3}$,$F_{trial4}$). This crossover step ensures 
the attacker can generate a diverse set of candidates to cover most of 
the DNN weight search space.
\vspace{-0.5em}
\paragraph{Selection Step.}
The selection step selects only the best candidate (i.e. winner with the highest fitness function value) between the trial vector set (\{$p_{trial},q_{trial}$\} with four trial vectors) and current candidate ($p_i,q_i$). Then, the rest four will be eliminated. 
The above discussed mutation, crossover and selection will repeat $z$ times to cover all candidates in the population set $S$. As a result, the initial randomly proposed $S$ will evolve over time to gradually approach 
the attacker-defined malicious objective. When $z$ times evolution is finished, the attacker could perform AWD attack at the winner (with the highest fitness function value in $S$) weight package during transmission. P-DES will check if the attack objective has been achieved. If yes, it stops. If not, it goes to the next iteration for a new round of attack iteration.

\begin{figure*}
    \centering
    \includegraphics[width=0.9\linewidth]{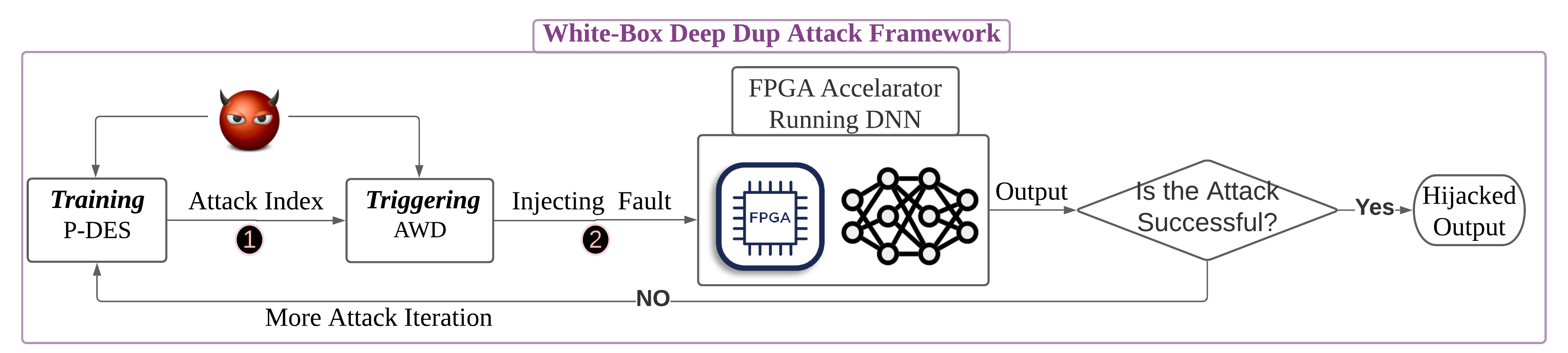}
    \caption{Overview of End-to-End \mlduplication attack framework integrating P-DES and AWD for White-Box attack}
    \label{fig:white_AWD}
    \vspace{-1.5em}
\end{figure*}

\vspace{-1em}
\subsection{End-to-End \mlduplication Attack}\label{sec:end-to-end}
\vspace{-0.5em}
This sub-section discusses the proposed end-to-end \mlduplication attack framework integrating \emph{training} software (i.e. searching) utilizing P-DES algorithm and hardware fault injection through AWD, i.e. fault \emph{triggering}. We also experimentally demonstrate the success of our end-to-end attack framework from the attacker's input end to the victim's output end for white-box and black-box attack. Note that, the fault injection reliability (i.e. fault injection success rate) and detection analysis will be discussed in detail in the experimental section \ref{sec:fault_rate} and \ref{sec_discussion}. The main mechanism of our \mlduplication attack framework could succeed even with real-world un-reliable hardware fault injection (i.e., with probability to succeed) is based on the fact that the vulnerable weight sets that our P-DES searching algorithm identifies are not static or unique, meaning the targeted attack index set could be progressively expanded based on real measured attack effect, for the same malicious objective. This is possible due to that deep learning model parameter training is a high dimension optimization process and many different fault injection combinations could lead to the same effect, which is also observed in prior works \cite{yao2020deephammer,Rakin_2019_ICCV,he2020defending}.
Thus, our proposed progressive evolutionary searching algorithm could take care of such fault injection uncertainty and randomness through redundant attack iterations to greatly improve the overall attack success rate, which is also experimentally validated in Sec.\ref{sec_black_box_attack_result} and \ref{sec_discussion}.


\vspace{-1em}
\subsubsection{White-Box Attack Framework}
\vspace{-0.5em}
\label{sec:white_frame}
\textbf{Training through P-DES.} As we discussed in the threat model, white-box attack assumes adversary knows all the details of target DNN model in victim FPGA, including architecture, weight values, gradients, weight package transmission over FPGA I/O protocol IP. 
.
As shown in Fig. \ref{fig:white_AWD}, knowing these execution details of 
the target DNN model, the adversarial can build an \textit{off-line simulator} (i.e. model replicate) to emulate the execution of target DNN in FPGA. 
Meanwhile, prior profiling should be conducted to estimate the fault injection success rate $f_p$ (84.84\% and 58.91\% for our measured RO and LRO based power plundering circuits), which will add randomness to the off-line simulated fault-injected DNN model and thus the fitness function evaluation (Eq.\ref{Eqn: fit}). Note that, this $f_p$ does not need to be very accurate. In general, smaller $f_p$ will force the progressive P-DES algorithm to generate 
a more redundant attack index to compensate 
for higher uncertainty of fault injection. More experiment results demonstrating the co-relation between $f_p$ and attack iterations are provided in Sec. \ref{sec_discussion} (Tab. \ref{tab:sweep}).
With the help of this off-line simulator, the P-DES searching algorithm will generate the attack index \circled{1}, i.e. model weight package index to be attacked during data communication.


\textbf{Triggering AWD.} 
In the next step \circled{2}, the P-DES generated attack index will be sent to our AWD triggering system to implement actual fault injection on those locations to achieve the defined malicious objective. More details of triggering system implementation are described in Sec.\ref{sec:triggering_sys}. To summarize, the attacker profiles the targeted DNN weight package indexes through 
the TDC sensor and embeds the received attack index from 
the last step into the attacking strategy file (Fig. \ref{fig:trigger_img}), which automatically triggers and controls the power-plundering circuits to implement the fault injection in the designed locations. After that, if the attack objective is not achieved (i.e., due to un-successful fault injection), the attacker will repeat the steps \circled{1} and \circled{2} to re-generate 
a more redundant attack index until successful. 

\vspace{-1.5em}
\subsubsection{Black-Box Attack Framework}\label{sec:bb_atk}
\vspace{-0.5em}

\begin{figure*}
    \centering
    \includegraphics[width=0.95\linewidth]{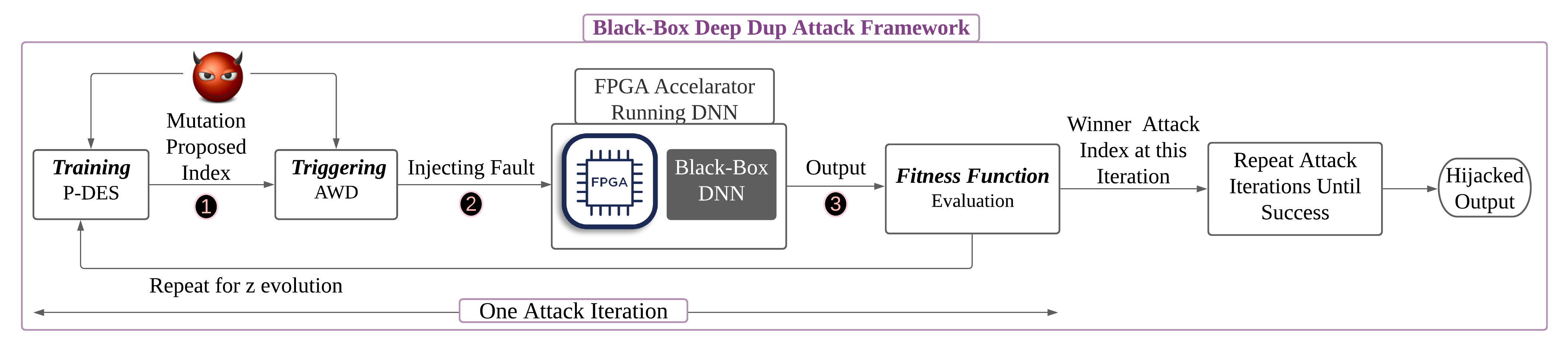}
    \caption{Overview of End-to-End \mlduplication attack framework integrating P-DES and AWD for Black-Box Attack.}
    \label{fig:black_AWD}
    \vspace{-1.5em}
\end{figure*}

Fig. \ref{fig:black_AWD} shows the overview of \mlduplication black-box attack framework.
Instead of constructing an off-line replicate to search vulnerable weights in white-box attack, in black-box attack, \mlduplication directly utilizes run-time victim DNN in target FPGA to evaluate the attack effectiveness (i.e. fitness function) of our searching algorithm P-DES proposed weight candidate in mutation step for every attack iteration. Thus, the un-reliable fault injection phenomenon is automatically considered and evaluated in the framework since the fitness function is directly evaluated in the victim FPGA using the real fault injection attack.

In this black-box setting, for every attack iteration, the attacker first utilizes the mutation function defined in our P-DES algorithm to propose a potential attack index candidate \circled{1}. Next, it will be sent to the AWD triggering component (Fig.\ref{fig:trigger_img}) to implement fault injection \circled{2} in current evolution. Therefore, the current DNN model in FPGA is executed based on the fault-injected model, where its DNN output \circled{3} will be read out by the attacker to be recorded as attack effectiveness (i.e. fitness function evaluation). Note that, during this process, the fault injection may succeed, or not. As for an attacker, since it is a black-box, he/she does not know about it. Only the victim DNN output response w.r.t. currently proposed attack index will be recorded and sent back to our P-DES software. Then, this step \circled{1}-\circled{2}-\circled{3} will repeat $z$ evolution times to select one winner attack index to finish the current attack iteration. After that, a new attack iteration will be started to find the next winner attack index until the defined attack objective is achieved.    

\textbf{ Modification of P-DES to adapt to Black-Box.}
For a black-box attack, the attacker can only access the input and output scores of the target DNN in victim tenant FPGA, with no knowledge of DNN architecture (i.e., in P-DES, $p$\ refers to \# of layers $\&$ $q$ refers to \# of weights at each layer) (details in section \ref{sec:p_des}). To adapt the P-DES algorithm to a black-box attack, instead of using architecture info of $p$ and $q$ (i.e., 2D vector), we will treat the whole network parameter to be unwrapped into a 1D vector $w$, where an attacker tries to identify each weight with one feature $\hat{p}$. Here, $\hat{p}$ denotes the weight index to be attacked after flattening and combining all $L$ layers weights sequentially. As we defined in the threat model section and AWD triggering section (sec.\ref{sec:triggering_sys}), this is feasible since the attacker knows which clock cycles are used to transmit DNN model weights, enabling an attacker to develop such a 1D weight index vector for the P-DES. This is the only modification needed for P-DES algorithm discussed in section \ref{sec:p_des} to adapt to black-box attack.


\textbf{Triggering AWD in Black-Box.} Most of the AWD triggering scheme (details in Sec.\ref{sec:triggering_sys} ) of black-box attack is similar to that in white-box (i.e., controlled by the attacking strategy file), except that it will be triggered much more frequently. The attacking strategy file (Fig. \ref{fig:trigger_img}) will be updated within every search evolution when it receives mutation proposed attack candidate, to trigger a new fault injection in the designated location for next fitness function evaluation in FPGA. $z$ evolution is needed for one attack iteration.

\textbf{Fitness Function Evaluation.} As discussed above, in a black-box setting, the attacker directly feeds a sample input into the FPGA to evaluate the fitness function in step \circled{3}. As the attacker can only access the output prediction from FPGA, he/she can compute the loss function using Eqn.\ref{eqt:unt} and Eqn.\ref{eqt:loss_T-BFA_yolo} for un-targeted and targeted attack, respectively. The above process \circled{1}-\circled{2}-\circled{3} continues for $z$ evolution times to select one winner candidate to finish one attack iteration. Then, it goes to the next iteration until the attack objective is achieved.


\vspace{-1.5em}
\section{Experimental Setup}
\vspace{-0.5em}
\vspace{-0.5em}
\subsection{Dataset and DNN Models }\label{sec:Dataset_DNN_arch}
\vspace{-0.75em}
In our experiment, we evaluate three classes of datasets. First, we use CIFAR-10 \cite{krizhevsky2010cifar} and ImageNet \cite{krizhevsky2012imagenet} for image classification tasks.  The other application is object detection where we evaluate the attack on the popular COCO \cite{lin2014microsoft} dataset. 

For CIFAR-10 dataset, we evaluate the attack against popular ResNet-20 \cite{he2016deep} and VGG-11 \cite{simonyan2014very} networks. We use the same pre-trained model with exact configuration as \cite{rakin2020tbt,he2020defending}. For ImageNet results, we evaluate our attack performance on MobileNetV2 \cite{sandler2018mobilenetv2}, ResNet-18 and ResNet-50 \cite{he2016deep} architectures. For MobileNetV2 and ResNet-18, we directly downloaded a pre-trained model from PyTorch Torchvision models \footnote{https://pytorch.org/docs/stable/torchvision/models.html} and perform an 8-bit post quantization same as previous attacks \cite{Rakin_2019_ICCV,rakin2020tbt}. For the ResNet-50, we use Xilinx 8-bit quantized weight trained on ImageNet from \cite{CHaiDNN}. The model we use to validate the YOLOv2 is the official weight \cite{YOLOV2_model}, trained by COCO \cite{lin2014microsoft} dataset, and we quantize \cite{shan2016dynamic} each weight value into 16-bits. Our code is also available publicly\footnote{https://github.com/ASU-ESIC-FAN-Lab/DEEPDUPA}.

 \vspace{-1em}
\subsection{FPGA Prototype Configurations} \label{sec:FPGA config}
\vspace{-0.5em}

\begin{figure}[htb!]
    \centering
    \includegraphics[width=0.6\linewidth]{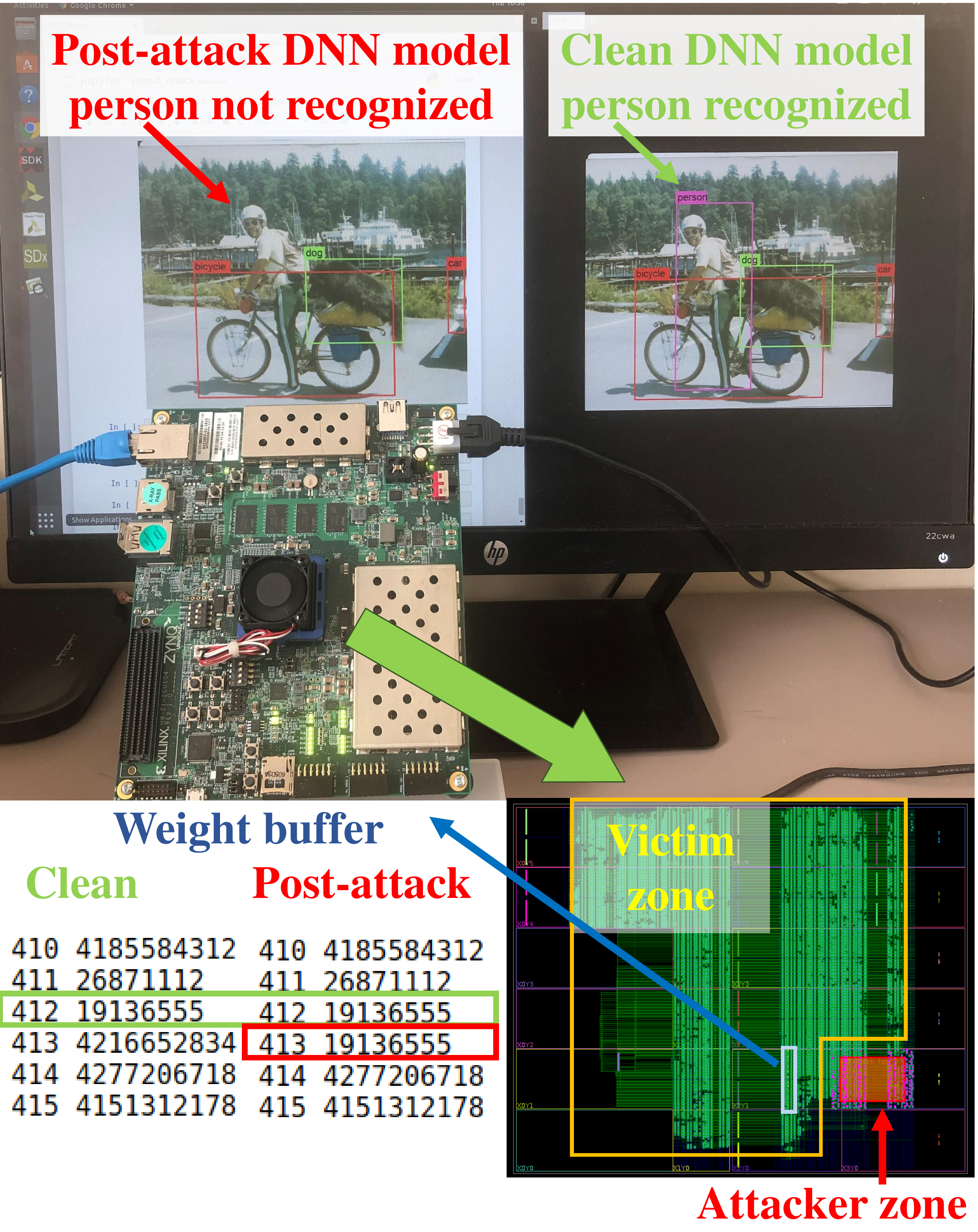}
    \caption{Experimental setup and results of \mlduplication black-box attack on YOLOv2, with `person' as target group. After attack, the fault-injected YoLov2 model fails to recognize the `person'.}
    \label{fig:setup}
    \vspace{-1em}
\end{figure}
To validate the real-world performance of \mlduplication,
we develop a multi-tenant FPGA prototype, using a ZCU104 FPGA evaluation kit with an ultra-scale plus family MPSoC chip, which has the same FPGA structure as these used in a commercial cloud server (e.g., AWS F1 instance), 
running the above discussed deep learning applications: image classification and object detection. 
The 8-bit quantized DNN models are deployed to our FPGA prototype through a high-level synthesis (HLS) tool, PYNQ frameworks, and CHaiDNN library from {Xilinx} \cite{CHaiDNN}. 
The experimental setup is shown in Fig. \ref{fig:setup}. For object detection (i.e. YOLOv2) FPGA implementation, multiple types of hardware accelerators (HAs) are used to compute different network layers, such as convolution layer, max-pooling layer, and reorganization layer. Specially, the region layer and data cascade are assigned to the ZYNQ's ARM core. For image recognition (e.g. ResNet-50) FPGA implementation, we follow the same design as 
the Xilinx mapping tool, which only implements the convolution accelerator in a light version (DietChai)\cite{CHaiDNN}. Without loss of generality, the FPGA configurations follow the official parameters \cite{yolov2github} and \cite{CHaiDNN}. Object detection network (i.e. YOLOv2) in FPGA execution frequency is 180MHz on 
Image recognition DNN network (e.g. ResNet-50) in FPGA execute frequency is 150MHz/300MHz, where the DSP uses a 300MHz clock source to increase the throughput and {for the other logic} we use a 150MHz clock.

To emulate a multi-tenant FPGA environment, we divide the FPGA resources into victim and attacker zones, respectively. The victim zone runs target DNN models, like YOLOv2 or ResNet-50, while the attacker zone mainly consists of malicious power-plundering circuits. 
Moreover, to limit the available resources of attacker, only 13.38\% of the overall FPGA resources are assigned for the power-plundering circuits.

\begin{table*}[tb!]
\centering
\caption{Summary of the White-Box Attack on CIFAR-10 and ImageNet Dataset. Here, $t_s$ denotes the target class which we randomly selected for each cases. The attack number is the best number out of three test rounds due to randomness.}
\label{tab:classification}
\scalebox{0.8}{
\begin{tabular}{@{}|c|c|c|c|cc|cccc|@{}}
\toprule
\multicolumn{4}{|c|}{ White-Box Attack on Image Recognition} & \multicolumn{2}{c|}{Un-Targeted Attack} & \multicolumn{4}{c|}{Targeted Attack} \\ \midrule
Dataset & Network & \# of Parameters & \begin{tabular}[c]{@{}c@{}}TA (\%)\end{tabular} & \begin{tabular}[c]{@{}c@{}} Post-Attack \\TA (\%) \end{tabular} & \begin{tabular}[c]{@{}c@{}}\# of \\ Attacks\end{tabular} & \begin{tabular}[c]{@{}c@{}}Post-Attack \\ TA (\%)\\ \end{tabular} & Target Class($t_s$) & \begin{tabular}[c]{@{}c@{}}ASR\\ (\%)\end{tabular} & \begin{tabular}[c]{@{}c@{}}\# of \\ Attacks\end{tabular} \\ \midrule
\multirow{2}{*}{CIFAR-10} & ResNet-20& 0.27 M & 90.77 & 10.92 & 28 & 21.63 & Bird & 99.2 & 14 \\ \cmidrule(l){2-10} 
 & VGG-11& 132 M & 90.38 & 10.94 & 77 & 23.68 & Horse & 98.6 & 63 \\ \midrule
\multirow{4}{*}{ImageNet} & MobileNetV2& 2.1 M& 70.79 & 0.19 & 1 & 8.93 & Lesser Panda & 100.0  & 1 \\
\cmidrule(l){2-10} 
 & ReNet-18& 11 M & 69.35 & 0.18 & 106 & 34.45 & Ostrich & 100.0 & 13 \\
\cmidrule(l){2-10} 
 & ReNet-50& 23 M & 72.97 & 0.19 & 175 & 30.57 & Ostrich & 100.0 & 20 \\ \bottomrule
\end{tabular}}
\vspace{-1.25em}
\end{table*}

\subsection{Evaluation Metric and Hyper-parameters} 
\vspace{-0.5em}
For classification application, we use \textit{Test Accuracy (TA)} as the evaluation metric. Test Accuracy is the percentage of samples correctly classified by the network. We denote the test accuracy after the attack as \emph{Post-Attack TA}. For a targeted attack, we use \textit{Attack Success Rate (ASR)} to evaluate the performance of 
the attack; ASR is the percentage of the target class samples miss-classified to an incorrect class after 
an attack. For the object detection application, we use \textit{Mean Average Precision (mAP)} as the evaluation metric that is the primary metric in the official COCO dataset challenge website\footnote{https://cocodataset.org/\#detection-eval}. In P-DES, the attack evolution ($z$) is set to (500/1000) (white-box) and 100 (black-box). In our un-targeted attack, we use a test batch containing 256/25 images for the CIFAR-10/ImageNet dataset. Our code is available publicly\footnote{https://github.com/ASU-ESIC-FAN-Lab/DEEPDUPA} with detailed hyper-parameters .

\section{Experimental Validation and Results}
\label{sec:results}
\vspace{-0.75em}


\subsection{{Measured Fault Injection Success Rate}}
\vspace{-0.5em}
\label{sec:injection_rate}
\label{sec:fault_rate}
As described in Fig. \ref{Fig:On-chip_buffer_status}, the AWD attack targets the weight transmission procedure, and the fault injection may not always succeed. However, it is infeasible to validate such fault injection success rate in our black-box attack model, in which the adversary has no access to the manipulated weight packages. To measure that, we design another experiment using an AXI4-based weight transmission with the same YOLOv2 setup, i.e., the same memory copy operation. We define the burst length of AXI4 as 256. 
The entire YOLOv2 int16 quantized weight (99496KB) needs 99496 bursts to finish the transmission for one input image inference. To avoid an FPGA system crash, we only trigger one attack at the middle transmission moment of a burst. 
To mimic the practical multi-tenant environment with the victim DNN model being executed simultaneously, we run a YOLOv2 in parallel.
The available power-plundering circuits are also the same as that in Sec. \ref{sec:FPGA config}. Using this experimental setup, we measured the success rates of fault injection by RO and LRO power-plundering circuits are 84.84\% and 58.91\%, respectively.

\vspace{-1.75em}

\paragraph{FPGA system crash avoidance.} It has been discussed in prior work \cite{gnad2017voltage} that a too-aggressive power attack (i.e., leveraging a large power-plundering circuit, or triggering it with unsuitable frequency and duty-cycle) will possibly cause an FPGA system crashes. In our case study, we limit the hardware resources available to the adversary. Additionally, to avoid such system crash, we apply two constraints on the triggering of AWD attacks: 1) A short activation period of each fault injection and 2) A large enough interval between any two consecutive fault injections. Specially, our experiment sets each fault injection period to 50 ns, from which we did not observe a crash of the FPGA setup. The attacking interval between each two consecutive fault injection is set to be longer than 600 ns, which is handled by our P-DES algorithm development, i.e., searching for target attack indexes with a certain distance in between.


\vspace{-1em}
\subsection{White-Box Attack Results}\label{sec:white_attack}
\vspace{-0.5em}
\paragraph{Image Classification Task.}
We evaluate the proposed \mlduplication white-box attack framework (in Fig. \ref{fig:white_AWD}) on two popular Image Classification datasets in Tab. \ref{tab:classification}. First, for CIFAR-10, our attack achieves close to the target random guess level accuracy (e.g., 10 \% for CIFAR-10) with only \emph{28} attack iterations (un-targeted) on ResNet-20. However, to deteriorate the test accuracy of VGG-11 to 10.94 \%  from 90.38 \%, \mlduplication requires \emph{77} attacks. Similarly, for targeted attack on CIFAR-10, the attacker requires only \emph{14} and \emph{63} attacks to achieve close to \emph{99.0} \% ASR on ResNet-20 and VGG-11 respectively. Clearly, VGG-11 is more robust to \mlduplication attack. We provide the detailed analysis of this phenomenon in sec.\ref{sec:defense}. 

For ImageNet dataset, our attack succeeds in degrading {the test accuracy of} MobileNetV2 to \emph{0.19} \% from \emph{70.79} \% with just one \emph{single} attack. Even for the targeted attack, it only requires one attack to achieve \emph{100} \% ASR in miss-classifying all \emph{Lesser Panda} images. Again, MobileNetV2 is also found to be extremely vulnerable by previous adversarial weight attack \cite{yao2020deephammer} as only a single bit memory error can cause catastrophic output performance. Nevertheless, MobileNet is an efficient and compact architecture ideal for mobile and edge computing platforms like FPGA \cite{wu2019high}. Thus the vulnerability of these compact architectures against \mlduplication raises a fair question of how secure are these DNN models in cloud FPGA? The answer from our \mlduplication attack is a \emph{big NO}. 
Our attack also succeeds in all ResNet families. Also, larger DNN models (e.g., ResNet-18 \& ResNet-50) shows better resistance to \mlduplication attack.

\begin{table}[tb!]
\centering
\caption{Black-Box targeted attack results for ImageNet.} 
\vspace{-0.5em}
\label{tab:resnet_bb}
\scalebox{0.9}{
\begin{tabular}{@{}ccccc@{}}
\toprule 
\multicolumn{5}{c}{\textbf{Black-Box Targeted Attack on ResNet-50 using RO cell}} \\ \midrule
\begin{tabular}[c]{@{}c@{}} ($t_s$)\end{tabular} & \begin{tabular}[c]{@{}c@{}}TA(\%) \end{tabular} &
\begin{tabular}[c]{@{}c@{}}Post-Attack TA(\%)\end{tabular} &
\begin{tabular}[c]{@{}c@{}} ASR  (\%)\end{tabular} &
\begin{tabular}[c]{@{}c@{}}\# of  Attacks\end{tabular} \\ 
\midrule
Ostrich  & 72.97 
& 46.96 & 100 & 26\\  
\bottomrule
\end{tabular}
}
\vspace{-1.5em}
\end{table}

\begin{table}[tb!]
\centering
\caption{Black-Box attack for object detection.}
\vspace{-0.6em}
\label{tab:yolo}
\scalebox{0.9}{
\begin{tabular}{@{}cccc@{}}
\toprule
\multicolumn{4}{c}{\textbf{Black-Box Un-Targeted Attack on YOLOv2 using RO cell}} \\ \midrule
\begin{tabular}[c]{@{}c@{}}Target Class ($t_s$)\end{tabular} & \begin{tabular}[c]{@{}c@{}}mAP \end{tabular} & \begin{tabular}[c]{@{}c@{}}Post- Attack mAP\end{tabular} & \begin{tabular}[c]{@{}c@{}}\# of  Attacks\end{tabular} \\ \midrule
All & 0.428 & 0.06 & 30 \\ \midrule
\multicolumn{4}{c}{\textbf{Black-Box Un-Targeted Attack on YOLOv2 using LRO cell}} \\ \midrule
\begin{tabular}[c]{@{}c@{}}Target Class ($t_s$)\end{tabular} & \begin{tabular}[c]{@{}c@{}}mAP \end{tabular} & \begin{tabular}[c]{@{}c@{}}Post- Attack mAP\end{tabular} & \begin{tabular}[c]{@{}c@{}}\# of  Attacks\end{tabular} \\ \midrule
All & 0.428 & 0.14 & 63 \\ \midrule
\multicolumn{4}{c}{\textbf{Black-Box Targeted Attack on YOLOv2 using RO cell}} \\ \midrule
\begin{tabular}[c]{@{}c@{}}Target Class ($t_s$)\end{tabular} & \begin{tabular}[c]{@{}c@{}}AP \end{tabular} & \begin{tabular}[c]{@{}c@{}}Post-Attack AP\end{tabular} & \begin{tabular}[c]{@{}c@{}}\# of  Attacks\end{tabular} \\ \midrule
Person &  0.6039 & 0.0507 & 20\\
Car &  0.5108 & 0.0621 & 18\\
Bowl &  0.3290 & 0.0348 & 15\\
Sandwich &  0.4063 & 0.0125 & 6\\ \bottomrule
\end{tabular}}
\vspace{-1.5em}
\label{tab:yolo_bb}
\end{table}

\vspace{-1em}
\subsection{Black-Box Attack Results}\label{sec_black_box_attack_result}
\vspace{-0.5em}

For proof of concept of our proposed \mlduplication black-box framework shown in Fig. \ref{fig:black_AWD}, in this section, we demonstrate and validate the black-box attack on Resnet-50 for image classification task and YOLOv2 for the object detection task. 
Specially, in our case study, we randomly pick the "ostrich" class in the Imagnet dataset as a target class for ResNet-50 and 4 target objects (i.e. Person, Car, Bowl and Sandwich) in the COCO dataset for YOLOv2. Other settings and performance metrics are the same as described in Sec. \ref{sec:white_attack}. Note that, all the black-box results are the actual measurement from our FPGA prototype. The \mlduplication black-box attack on ResNet-50 are successful and results are reported in Tab. \ref{tab:resnet_bb}. It can be seen that only 26 attacks are needed to attack the ``ostrich'' with 100 \% ASR. Similarly, \mlduplication black-box un-targeted and targeted attacks on YOLOv2, with both RO and LRO cells, are also successful, as reported in Tab. \ref{tab:yolo_bb}. It can be seen that the post-attack average precision (AP) is significantly degraded after less than 20 attacks. For example, only 6 attacks are needed to decrease the AP of sandwich class from 0.4063 to 0.0125.


\vspace{-1.4em}
\subsection{Comparison to Other Methods}
\vspace{-0.5em}
Previously, very few adversarial weight attack works have been successful in attacking DNN model parameters to cause complete malfunction at the output \cite{liu2017fault,hong2019terminal}. Thus we only compare with the most recent and successful adversarial bit-flip (BFA) based weight attack \cite{Rakin_2019_ICCV,yao2020deephammer}, which uses a gradient-based search algorithm to degrade DNN performance in a white-box setting. We also compare our search algorithm (P-DES) to a random AWD attack.
\vspace{-0.25em}
\begin{table}[ht]
\centering
\caption{Comparison of \mlduplication with \emph{random} AWD attack and row-hammer based (\emph{BFA} \cite{Rakin_2019_ICCV,yao2020deephammer}) attack. All the results are presented for 8-bit quantized VGG-11 model \cite{Rakin_2019_ICCV}.}
\vspace{-0.5em}
\label{tab:cmp}
\scalebox{0.75}{
\begin{tabular}{@{}ccccc@{}}
\toprule
Method & \begin{tabular}[c]{@{}c@{}}Threat\\ Model\end{tabular} & \begin{tabular}[c]{@{}c@{}}TA \\  (\%)\end{tabular} & \begin{tabular}[c]{@{}c@{}} Post-Attack TA  (\%)\end{tabular} & \begin{tabular}[c]{@{}c@{}}\# of Attacks\end{tabular} \\ \midrule
Random & \begin{tabular}[c]{@{}c@{}}Black Box\end{tabular} & 90.23 & 90.04 & 100 \\
BFA \cite{yao2020deephammer} & \begin{tabular}[c]{@{}c@{}}White Box\end{tabular} & 90.23 & 10.8 & 28 \\
\begin{tabular}[c]{@{}c@{}} \mlduplication \end{tabular} & \begin{tabular}[c]{@{}c@{}}Black \& White Box\end{tabular} & 90.23 & 10.94 & 77 \\ \bottomrule
\end{tabular}}
\vspace{-1.25em}
\end{table}

As shown in both Tab. \ref{tab:cmp} , only \emph{77} AWD attack iterations can degrade the accuracy of VGG-11 to \emph{10.87} \% while randomly performing \emph{100} AWD attacks, cannot even degrade the model accuracy beyond \emph{90} \%. On the other hand, a BFA attack \cite{yao2020deephammer} using row-hammer based memory fault injection technique, requires only 28 attacks (i.e. memory bit-flips) to achieve the same un-targeted attack success (i.e., $\sim$ 10 \% TA). However, BFA attack is only successful for white-box setting, not black-box.

\vspace{-1.5em}
\subsection{Discussion} \label{sec_discussion}
\vspace{-0.5em}



\paragraph{Attack efficiency w.r.t. fault injection success rate.}
As described in section \ref{sec:fault_rate}, we used two different power plundering circuits, i.e., RO and LRO for fault injection. In our experiments,  we measured 84.84\% and 58.91\% fault injection success rates for RO and LRO, respectively. In practical attack, this number may vary due to the attack budget (i.e., frequency, resource, etc.). In order to validate our \mlduplication attack framework will succeed in different fault injection success rates, we incorporate the fault success rate as a probabilistic parameter in our off-line simulator as discussed in section \ref{sec:white_frame}. Note that, for black-box attack, our direct evaluation of fitness function in the FPGA accelerator already considers and compensates for the failed fault iteration. 
The experimental results are shown in Tab.\ref{tab:sweep}. We observe that our \mlduplication attack framework could still succeed at  very low fault injection success rate (i.e., 40 \%), but requiring more number of attack iterations (i.e. higher redundancy as explained in sec.\ref{sec:end-to-end}).

\vspace{-0.5em}
\begin{table}[ht]
\centering
\caption{Attack efficiency v.s. fault injection success rate ($f_p$). Reporting  \# of attack iterations (i.e., mean $\pm$ std. for three runs) required to achieve 99.0 \% ASR (targeted attack) or 11.0 \% test accuracy (un-targeted attack).}
\vspace{-0.5em}
\label{tab:sweep}
\scalebox{0.8}{
\begin{tabular}{@{}ccccc@{}}
\toprule
Model & Type & 40 \% & 60 \% & 80 \% \\ \midrule
\multirow{2}{*}{ResNet-20} & Un-Targeted & 95.3 $\pm$ 37.3 & 88 $\pm$ 66.5 & 76.6 $\pm$ 13.8 \\
 & Targeted & 39 $\pm$ 7.8 & 23.3 $\pm$ 4.3  & 23.8 $\pm$ 6.8  \\ \midrule
\multirow{2}{*}{VGG-11} & Un-Targeted & 195.3 $\pm$ 39.1 & 95.6 $\pm$ 14.1 & 98.9 $\pm$ 1.9 \\
 & Targeted & 114 $\pm$ 32 & 88.6 $\pm$ 34.4 & 62.6 $\pm$ 2.6  \\ \bottomrule
\end{tabular}}
\vspace{-1.5em}
\end{table}

\begin{figure}[tb!]
    \centering
    \includegraphics[width=1\linewidth]{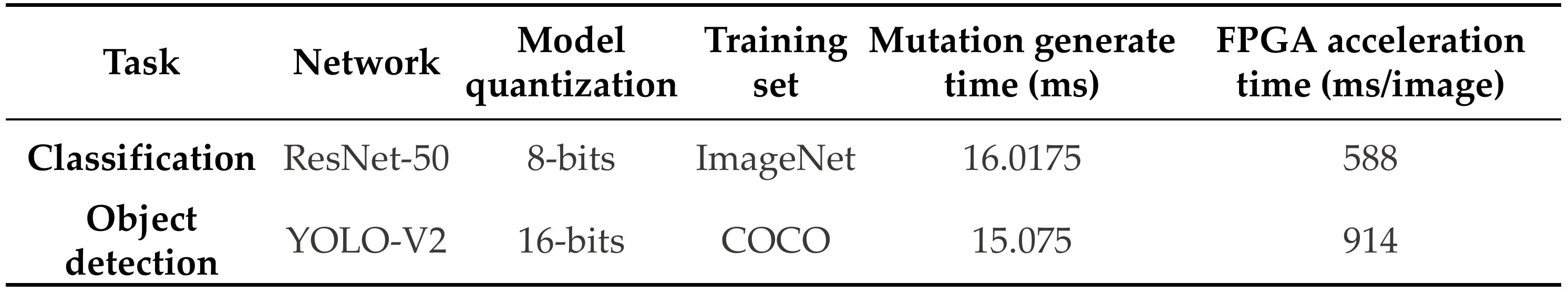}
    \caption{Black-Box attack time cost analysis with $z =100$. FPGA acceleration (i.e., fitness function evaluation) time and mutation generation time are reported.} 
    \label{fig:Per_eval}
    \vspace{-1.75em}
\end{figure}

\vspace{-0.5em}
\paragraph{Attack Time Cost.} The execution time of one searching iteration of our proposed P-DES algorithm is constant for a fixed $z$, regardless of DNN model size. The overall searching time is proportional to the number of evolution ($z$). For \mlduplication white-box attack, the P-DES algorithm is executed offline, and the AWD attack is only executed when the attack index is generated. Note that, the hardware AWD attack incurs no time cost, as it runs in parallel with the victim DNN model. For \mlduplication black-box attack, two main time cost includes mutation generation (proportional to $z$) and FPGA fitness function evaluation (proportional to DNN acceleration performance/latency in FPGA).  
In Fig. \ref{fig:Per_eval}, we report the average time cost of the proposed 4 mutation strategies executed in the PS of our FPGA prototype. Additionally, we also report the DNN execution time in FPGA, which is determined by the corresponding DNN model size, architecture, optimization method, and available FPGA hardware resources. It is easy to observe that our P-DES mutation generation only consumes trivial time compared to DNN execution time in FPGA, which is the bottleneck in black-box attack.


\vspace{-1.25em}
\section{Potential Defense Analysis}
\label{sec:defense}

\vspace{-1em}
\paragraph{Increasing Model Redundancy.}
Several prior works have demonstrated that increasing model redundancy (i.e., DNN size/channel width) \cite{he2020defending,li2019d2nn} can be a potential defense against model fault attack. Our evaluation of \mlduplication attack in the previous section also indicates the 
correlation between network capacity (i.e., \# of model parameters) and model robustness (\# of attacks required). As the ImageNet dataset section depicts in Tab. \ref{tab:classification}, as the network size increases from ResNet-18 to ResNet-50, the number of attacks required to achieve 100 \% ASR increases correspondingly. We observe the same trend for CIFAR-10 models where VGG-11 (i.e., dense model) requires a higher number of attacks than ResNet-20 (i.e., compact model).
\vspace{-1em}
\begin{table}[ht]
\centering
\caption{Attack efficiency after increasing the model size of ResNet-20 and VGG-11 model by 4 (i.e., increasing each input and output channel size by 2).}
\vspace{-0.5em}
\label{tab:model_size}
\scalebox{0.85}{
\begin{tabular}{ccc}
\hline
Method & ASR(\%) & \# of  Attacks \\ \hline
ResNet-20 (\emph{Baseline}) & 99.6 & 14 \\
ResNet-20 \emph{$\times$ 4} & 99.6 & \emph{21} \\ \hline
VGG-11 (\emph{Baseline}) & 98.6 & 63 \\
VGG-11 \emph{$\times$ 4} & 98.2 & \emph{84}  \\ \hline
\end{tabular}}
\vspace{-0.75em}
\end{table}

In Tab. \ref{tab:model_size}, we run an experiment to validate the relation between \mlduplication attack efficiency and network model size. First, we multiply the input and output channel of the baseline model by 2 to generate ResNet-20 ($\times$ 4) and VGG-11 ($\times$ 4) models with 4 $\times$ larger capacity. For both ResNet-20 and VGG-11, the number of attacks required to achieve similar ASR increases with increasing model capacity (Tab. \ref{tab:model_size}). 
To conclude, one possible direction to improve the DNN model's resistance to the \mlduplication attack is to use a dense model with a larger redundancy.

\vspace{-1.5em}
\paragraph{Protecting Critical Layers.}
Another possible defense direction is to protect the critical layers that are more sensitive. Prior works \cite{8554308} have proposed selective hardening to defend against weight faults by selectively protecting more sensitive layers. It is interesting to note that our experimental observation also shows that 80 \% of the searched vulnerable weights are within the first two layers and the last layer for ResNet-20. Following this observation, in Tab. \ref{tab:protected}, we run our attack by securing these three sensitive layers (\emph{ResNet-20 (Protected)}). A straightforward way to secure layer weights from \mlduplication would be to store them on-chip (i.e., no need for off-chip data transfer). Note that, a defender can not store an entire DNN model on-chip due to limited on-chip memory and typically large DNN model size for cloud computing. Nevertheless, as shown in Tab. \ref{tab:protected}, our \mlduplication still manages to succeed with $\sim$ \emph{2} $\times$ additional rounds of attack on the protected ResNet-20 model.
Similarly for VGG-11, our \mlduplication attack still successfully achieves $\sim$ 99.0 \% ASR even after securing some critical DNN layers from fault attacks.

\vspace{-0.75em}
\begin{table}[ht]
\centering
\caption{\mlduplication attack performance after protecting or securing some critical DNN layers}
\vspace{-0.5em}
\label{tab:protected}
\scalebox{0.85}{
\begin{tabular}{ccc}
\hline
Method & ASR(\%) & \# of  Attacks \\ \hline
ResNet-20 (\emph{Baseline}) & 99.6 & 14 \\
ResNet-20 (\emph{Protected}) & 99.2 & \emph{29} \\
VGG-11 (\emph{Baseline}) & 98.6 & 63 \\
VGG-11(\emph{Protected}) & 98.2 & 141 \emph{}\\ \hline
\end{tabular}}
\vspace{-1.5em}
\end{table}


\paragraph{Obfuscation through Weight Package Randomization.}
In our \mlduplication attack, 
the P-DES algorithm relies on the sequence (e.g., index) of the weight packages being transferred between the on-chip buffer and off-chip memory. In this section, we discuss the possibility of defending our attack by introducing random weight package transmission as an obfuscation scheme. In Tab. \ref{tab:randomization}, we first perform an experiment with shuffling of the weights in a pre-defined sequence before transmitting them. The results show that pre-defined shuffling order of the wights has almost no effect on the attack efficacy.
\vspace{-1.5em}
\begin{table}[ht]
\centering
\caption{Weight package randomization as obfuscation. \textbf{Pre-defined Shuffle} : Shuffling the weight packages in a pre-defined order before transmission. \textbf{Random Shuffle} : Shuffling the weight packages every time using a random function before transmission.}
\vspace{-0.5em}
\label{tab:randomization}
\scalebox{0.85}{
\begin{tabular}{@{}cccc@{}}
\toprule
Method & \begin{tabular}[c]{@{}c@{}}TA (\%)\end{tabular} & \begin{tabular}[c]{@{}c@{}}Post-Attack\\ TA (\%)\end{tabular} & \begin{tabular}[c]{@{}c@{}}\# of \\ Attacks\end{tabular} \\ \midrule
Random Attack & 90.77  &  87.9 & 180\\
ResNet-20 Baseline & 90.77 & 10.94  & 28 \\
Pre-defined Shuffle & 90.77  & 11.0 & 26  \\
Random Shuffle& 90.77  & 53.3 & 180\\
\bottomrule
\end{tabular}}
\vspace{-1em}
\end{table}

Next, we discuss the case with shuffling the weight package for every transmission round as a very strong obfuscation. The effect of such a strong obfuscation scheme can have three possible implications. First, a randomly shuffled weight transmission will fail to defend our attack in a white-box setting as the attacker has full knowledge of the DNN and data transmission scheme. Second, in a black-box setting, as shown in Tab. \ref{tab:randomization}, this defense will greatly limit the efficacy of our attack, requiring 
a larger amount of attack iterations (e.g., 180) to degrade the accuracy to 53.3 \%. But the attack remains more successful than a random AWD attack with no searching algorithm. It aligns with the recent work of adversarial input attack \cite{athalye2018obfuscated}, where the authors argue that obfuscation based on 
an under-lying random function as defense may not completely defend a progressive adversarial attack. Given a large amount of model query, the progressive evolutionary 
algorithm-based attack (i.e. our case) could estimate the effect and distribution of the randomness to improve the attack efficacy in comparison to a random attack. Moreover, randomly shuffling data transmission every time would require additional header information to synchronize the sequence of weights at the receiver end. A recent work in  \cite{zhou2016high} has demonstrated random shuffling may cost 
up to 9 $\times$ energy in-efficiency and 3.7 $\times$ lesser amount of throughput. Thus, an effective defense scheme will always come at the expense of additional (i.e., memory, speed \& power) overhead. 
\vspace{-1.5em}
\paragraph{Power-based side-channel analysis to detect \mlduplication.}

Here we discuss the feasibility of using power-based side-channel analysis to detect \mlduplication. The success of such detection should rely on the ability to distinguish between these two cases: 1) \textit{Normal case}: two benign users execute their applications simultaneously, and 2) \textit{Attack case}: two users share the FPGA resources, where one of them apply \mlduplication to attack the other one. Since it is impractical to measure the real-time power trace in a cloud-FPGA with an oscilloscope, an on-chip power sensor (e.g., TDC sensor) will be the only option. As shown in Fig.\ref{fig:trigger_img}, similar as AWD attack, our measured power trace of a benign user (e.g., YOLOv2) also incurs large power glitches.
More importantly, we did not observe any AWD attack power glitch has a larger magnitude than that of benign user-YOLOV2. Instead, it is smaller for most of the time. 
Therefore, the glitches caused by AWD will be easily obfuscated. Further, it is difficult to distinguish AWD power glitches in the following practical scenarios: i) Most cloud-FPGA users prefer to run compute-intensive applications, which generates many power glitches; ii) When triggered, each fault injection by AWD only lasts for a short time period (e.g., 50ns) and is disabled for most of the time; iii) Faults are only injected at attacker's will, i.e., without a fixed pattern to check. In other words, it is of different challenges to use such power-based side-channel analysis for defense and attack, i.e., the defender should acquire ultra-high-resolution side-channel information to identify the malicious power glitches from the noisy power background by the compute-intensive application, e.g., the DNN execution; while the attacker only needs to identify the temporal range for the DNN weight transmission. More severely, an attacker may even choose to inject faults in a more stealthy manner, i.e., while the victim DNN model itself is generating lots of power glitches, to exacerbate the overall voltage drop \cite{luo2020stealthy}. Therefore, we argue that it is extremely difficult, if not impossible, to detect the proposed \mlduplication attacks with power anomaly in a multi-tenant FPGA. 

\vspace{-1.5em}
\section{Conclusion}
\vspace{-1em}
In this work, we study the security of DNN acceleration in multi-tenant FPGA. For the first time, we exploit this novel attack surface where the victim and the attacker share the same FPGA hardware sources. Our proposed \mlduplication attack framework is validated with a multi-tenant FPGA prototype, as well as some popular DNN architectures and datasets. 
The experimental results demonstrate that the proposed attack framework can completely deplete DNN inference performance to as low as random guess or attack a specific target class of inputs. It is worth mentioning that our attack succeeds even assuming the attacker has no knowledge about the DNN inference running in FPGA, i.e. black-box attack. A malicious tenant with such limited knowledge can implement both targeted and un-targeted malicious objectives to cause havoc for a victim user. 
Finally, we envision that the proposed attack and defense methodologies will bring more awareness to the security of deep learning applications in the modern cloud-FPGA platforms. 

\noindent\textbf{Acknowledgement:} 
The authors thank the designated shepherd (Dr. Nele Mentens) for her guidance, and the anonymous reviewers for their valuable feedback. This work is supported in part by the National Science Foundation under Grant No.2019548 and No.2043183.
\vspace{-1em}

\bibliographystyle{unsrt}

\bibliography{references}

\end{document}